\documentclass[12pt]{article}
\usepackage{authblk}
\usepackage{geometry}
\usepackage{amsmath}
\usepackage{amssymb}
\usepackage{amsthm}
\usepackage{mathrsfs}
\usepackage{subcaption}
\usepackage{customcommands}
\usepackage{bm}
\usepackage{Nettastyle}
\usepackage{physics} 
\usepackage{tensor}
\usepackage{tcolorbox}
\usepackage{multicol, makecell, longtable} %Better columns

\usepackage[normalem]{ulem}

\preprint{}

\title{Holographic Time Crystals vs Penrose}

\author{\AA{}smund Folkestad}
\emailAdd{afolkest@ucdavis.edu}
\affiliation{Center for Quantum Mathematics and Physics (QMAP), \\
Department of Physics \& Astronomy, University of California, Davis, CA 95616, USA}

\abstract{
In the large$-N$ limit, no known no-go theorem rules out thermal time crystals
that spontaneously break continuous time translation, unlike in the large
volume limit.  If thermal time crystals exist in holographic CFTs, they would
correspond to ensemble-dominating black holes with eternally time-varying
exterior geometries.  We point out that recent work on a conjectured non-linear
instability of slowly rotating Kerr-AdS$_4$ produced viable candidates for such
states.  Then we show that the existence of holographic microcanonical time
crystals would imply violations of the AdS Penrose inequality (PI). We proceed
to look for violations of the PI in spherical symmetry, working with
Einstein-scalar gravity with the most general possible boundary conditions
compatible with boundary conformal invariance.  We derive a set of ODEs for
maximally PI-violating initial data. Solving these numerically, we find strong
evidence that in the particular case of spherical symmetry, the PI holds iff the
positive mass theorem (PMT) holds. This suggests that holographic CFT$_3$ time crystals can
 only possibly exist at non-zero angular momentum, at least in the absence of electric
charge. We also discover neutral hairy black holes in a consistent truncation of
M-theory that has a PMT and boundary conditions respecting conformal invariance, disproving an
existing no-hair conjecture.  Finally, we show that previous PI-violating
solutions by the author all existed in theories where the PMT 
is violated.  Unfortunately, our results imply that there currently are no known
examples where the PI functions as a non-trivial Swampland constraint.
}

\begin{document}

\maketitle
\newpage

\section{Introduction}
In \cite{Wil12} a remarkable new phase of matter was proposed to exist: a time
crystal, defined by spontaneous breaking of time-translation symmetry. However,
the proposal was controversial \cite{Bru12, Wil13, LiGon12,
Bru13,TonZhe13,Bru13b}, and a later no-go theorem \cite{WatOsh14} showed under
quite general assumptions that
time crystals cannot exist, neither in the ground state nor at finite
temperature (see also \cite{Bru13b}). Subsequently, interest shifted to periodically driven systems, where
instead a discrete time translation symmetry undergoes spontaneous symmetry
breaking (SSB)--so-called Floquet time crystals \cite{Sac14,ElsBau16,Els16,Zha17,KheLaz16,Cho17,YaoPot17,KheMoe19}.  

However, the no-go theorem of \cite{WatOsh14}
relies on taking the infinite volume limit. While SSB can only ever exist in an
appropriate thermodynamic limit, other parameters than the volume can be blown
up to achieve SSB.  Thus, we can wonder whether time crystals exist in
other limits, such as the large$-N$ limit \cite{tHo74} that often appears in QFT
and the
AdS/CFT correspondence \cite{Mal97}. Here $N$ corresponds to a measure of the local number of
degrees of freedom, such as the rank of a gauge group for $\mathcal{N}=4$ super
Yang-Mills, or the central charge in two-dimensional CFT. 

In this paper, we point out that existing gravitational results in AdS
\cite{FigRos23} are suggestive of holographic large-$N$ time crystals possibly existing
at finite temperature. Then, we show that studying the so-called Penrose
inequality (PI) is a fruitful arena for getting closer to settling the matter.
In particular, the existence of microcanonical holographic time crystals implies
violation of the AdS-PI.\footnote{Strictly speaking, the entropy needs to scale
with $1/G_N \sim N^{a > 0}$ for this to be true.} After that, we restrict to four bulk dimensions,
spherical symmetry, and gravity coupled to a real scalar field,
whose CFT$_3$ dual is in a uniform state on a spatial $S^2$. There we carry out
a comprehensive numerical study of the PI. We derive a set of ODEs
for ``maximally PI-violating'' initial data (subject to a mild assumption),
and solve these numerically for Einstein-scalar theories with the most
general boundary conditions compatible with boundary conformal
invariance. This includes the regime of boundary scaling dimensions $\Delta \in
\left(\frac{ 1 }{ 2 }, \frac{ 3 }{ 2 }\right)$, where the naive ADM mass diverges, and where the
(finite) Hamiltonian generator gets corrections from scalar fields and even scalar
self-interaction coupling constants. 

For the theories in question, we find evidence that the spherically symmetric
AdS-PI is true if and only if the positive mass theorem (PMT) holds.  This
suggests that finite-volume holographic CFT$_{3}$ time crystals without electrical charge, if they were to exist,
only exist at non-zero angular momentum $J$.  Searching for violations of the
PI, we also come across neutral hairy black holes in a consistent
truncation of M-theory \cite{GauSon09} (see \eqref{eq:th2} for the scalar
potential).  While these black holes do not dominate the microcanonical
ensemble, they exist in a theory with a PMT and with conformally invariant
boundary conditions, which to our knowledge provides the first counterexample to
the no-hair conjecture of \cite{Her06}.

We also find that
previous PI-violating examples by the author \cite{Fol22} arose in
theories with lower unbounded energy, unlike what was believed based on the
numerical evidence presented there. Unfortunately, this means that there
currently are no known examples where the PI serves as a
non-trivial Swampland \cite{Vaf05,OogVaf06} constraint, unlike what was proposed
in \cite{Fol22}. Furthermore, as discussed below, before we can
know if it can possibly serve as a Swampland constraint at all, the issue of
time crystals must be settled.

The plan for the paper is as follows. In Sec.~\ref{sec:timecrystal} we review
how holographic time crystals are plausible in light of existing results.
Importantly, existing work already provide candidate
solutions. Then in Secs.~\ref{sec:PI1} and \ref{sec:PI2} we review the PI, how
all existing derivations of the PI take as an assumption that time crystals do not exist, and
explain how holographic time crystals with entropy of order $1/G_N$ would imply violation of the PI.
Then we review Einstein-scalar gravity with so-called ``designer gravity'' boundary
conditions, and associated positive mass theorems in Secs. \ref{sec:designer1} and
\ref{sec:designer2}. In Sec.~\ref{sec:newODE}  we derive an ODE system for time-symmetric initial datasets that have
minimal mass at fixed entropy, and we argue why it is reasonable to consider
time-symmetric initial data. In Sec.~\ref{sec:results} we present our numerical
results, including novel hairy black holes in Sec.~\ref{sec:deltaone}. We
conclude with a summary and discussion of future directions.

\section{The Penrose Inequality, Time Crystals, and Holography}
\subsection{Candidate solutions to holographic time
crystals}\label{sec:timecrystal}
Let us now imagine what the putative holographic dual to a thermal time crystal would look
like.\footnote{We focus on thermal states only, since we have no hints that
ground state time crystals could exist in the large-$N$ limit. Ground state time
crystals seem highly unlikely.} 
It is well understood that in the large$-N$ limit and at strong coupling, thermal ensembles
in the CFT usually are dual to classical eternal black holes
\cite{Mal01}.\footnote{Provided entropy is order $1/G_N$. 
Otherwise we can instead have thermal gasses of particles in non-black hole backgrounds \cite{HawPag83}. } 
Here we will always work in the microcanonical ensemble at fixed energy $E$ and
angular momentum $J$, so that the bulk states
that dominate the ensemble are the ones with maximal (HRT-)entropy \cite{Mar18}.
Now, SSB of time translation implies that the
black hole exterior geometry evolves with time forever. Given the dissipative
tendencies of black holes, naive intuition suggests that these eternally
non-stationary geometries should not exist. However, when the cosmological
constant is negative, black holes with eternal time-dependent exteriors were
in fact conjectured to exist almost 20 years ago \cite{KunLuc06}, and quite
recently they have been found to exist
\cite{DiaHor11,DiaSan15,IshMur18,IshMur21,Ish21}, proving our naive intuition
wrong. 

Let us now remind ourselves of a few facts about
Kerr-AdS$_4$. Kerr-AdS$_4$ is the simplest family of four-dimensional rotating
black holes with a negative
cosmological constant. The conformal boundary of these black holes is
$\mathbb{R}\times S^2$ with conformal structure represented by $-\dd
t^2+\dd\Omega^2$. Thus, if we worked with a holographic CFT$_3$ on this
geometry, it would be natural to expect that thermal states with angular
momentum are dual to Kerr-AdS$_4$. However, Kerr-AdS$_4$ has a superradiant
instability \cite{HawRea00,CarDia04,CarDia13} when the angular frequency
$\Omega_{H}$ exceeds
the inverse AdS radius: $\Omega_H > 1/L$. Superradiance \cite{Zel71,Sta73,BriCar15} is the effect
where there exist modes that reflect off the black hole with increased amplitude,
stealing some of the black hole's energy and angular momentum. Given
that the AdS conformal infinity acts as a reflecting boundary, these amplified
modes in turn bounce off it and fall back into the bulk in finite time. And so
this repeats, leading to an AdS realization of the black hole bomb \cite{PreTeu72}.
Eventually backreaction gets strong, and a longstanding question has
been what the endpoint of this instability is.

Ref. \cite{DiaSan15} constructed an interesting class of new black holes known as
\textit{black resonators}. These are vacuum
black holes with the same asymptotically AdS$_4$ boundary conditions as
Kerr-AdS$_4$. They are however neither axisymmetric nor stationary--they only have
a single Killing vector $\mathcal{K}=\partial_t + \Omega_H \partial_{\phi}$ with
$\Omega_H > 1/L$. As a consequence, $\mathcal{K}$ is spacelike asymptotically,
and the solutions are periodic in time rather than having a continuous
time-translation symmetry. Their horizonless limit are the AdS geons
\cite{DiaHor11b,HorSan14,MarFod17}.
These solutions furthermore have higher entropy than Kerr-AdS$_4$.
Nevertheless, also these
solutions are unstable. This follows from the general theorem of \cite{GreHol15}:
any AdS black hole with a somewhere spacelike Killing vector in the domain of
outer communication (causal wedge) is
unstable. 

Full non-linear numerical time-evolution of perturbed
Kerr-AdS$_4$ in the superradiantly unstable regime has been carried out by
\cite{CheLow18,Che21}. They found that Kerr-AdS$_4$ first evolves to a state
close to a black resonator. Then after a while it further evolves to a state
close to a \textit{multi-oscillating black resonator} \cite{IshMur21}.  This
state appears stable within the timescale simulated in \cite{Che21}, but it
could in principle be unstable over longer timescales.  In fact, it likely is
unstable, given that even more entropic solutions dubbed \textit{Grey Galaxies}
were constructed in \cite{KimKun23} and conjectured to be the final endpoint of
the instability of rapidly rotating Kerr-AdS$_4$.\footnote{In the large$-N$
limit ($G_N\rightarrow 0$), the CFT stress tensor becomes the sum of the usual
Kerr-AdS$_4$ contribution together with a delta-function contribution localized
around the equator of $S^2$. Grey Galaxies dominate the microcanonical ensemble,
but not the canonical ensemble.} These are $\Omega_{H}=1/L$ Kerr-AdS$_4$ black
holes surrounded by a large thin disk of spinning thermal gas, formed by matter
angular momentum modes with $\ell=m$ that have $O(1)$ occupation number all the
way up to $\ell = \mathcal{O}(1/\sqrt{G_N})$, implying that the disk stretches
out to radii of order $r \sim \mathcal{O}(1/G_N^{1/4})$.  Since these states are
stationary, the regime $\Omega_{H}>1$ does not correspond to a time crystal
(assuming no other even more entropic solutions exist). However, it is
interesting to note that it must take a time of order $e^{\#/\sqrt{G_N}}$
to evolve to a Grey Galaxy starting with Kerr-AdS$_4$ or a black
(multi-)resonator \cite{KimKun23}, so sufficiently rapidly spinning CFT$_{3}$
pure states\footnote{More precisely, states that look approximately classical
at $t=0$, i.e. having no bulk features scaling with $N$ to a positive power so
that the strict large-$N$ limit breaks down.} seem to never equilibrate in the
large-$N$ limit, even when their holographic duals are described by black holes
in Einstein gravity. Given that these large$-N$ dynamics are possible, perhaps a
rotating CFT$_3$ is still a good place to look for large$-N$ time crystals?  Or,
to beat less around the bush: what happens in the slowly rotating regime
($\Omega_H < 1/L$)?

It turns out that Kerr-AdS$_4$ for $\Omega_{H} < 1$ is linearly stable
\cite{HawRea00}. Nevertheless, it has been rigorously proven \cite{HolSmu11,HolWar14} that
scalar fields on a slowly rotating Kerr-AdS$_4$ background decay extremely
slowly (inverse logarithmically), leading mathematicians to conjecture
\cite{HolSmu11} that slowly rotating Kerr-AdS$_4$ is non-linearly unstable. This conjecture was recently investigated
in an interesting paper by Figueras and Rossi \cite{FigRos23}, who carried out full non-linear numerical
time-evolution of a perturbed Kerr-AdS$_4$ black hole with
$\Omega_{H} \approx 0.7/L$.\footnote{They included a massless scalar field for
convenience, although they expect the result to hold true in vacuum gravity as
well. Through the simulation, the scalar field shows evidence of decaying to
zero, meaning that end state is approximately vacuum.} They found evidence for the non-linear
instability conjecture. The perturbed black hole did not settle down to a member
of the Kerr-AdS$_4$ family. Instead, it ``settled down'' to a
non-stationary, non-axisymmetric black hole characterized by oscillations with two
different time scales. In the CFT this results in an energy
density with time-dependent $\ell=m\neq0$ modes oscillating with two
time scales, showing no signs of decay over the time of the simulation, which
lasted for $\Delta t \sim 200L$ -- roughly two orders of
magnitude larger than the AdS light crossing time and the timescale set by the
mass of the black hole. If their final state is indeed stable, then the most
obvious interpretation appears to be a genuine thermal large-$N$ time crystal in the microcanonical
ensemble.

Of course, from the numerics alone one cannot
rule out further dynamics over longer timescales, such as the evolution into a new
type of stable slowly rotating black resonator or a gradual conversion of energy
into higher and higher $\ell=m$ modes, which likely happens in the quickly
rotating case.\footnote{As explained in \cite{FigRos23},
$\ell=m$ modes dominate over $|m|<\ell$ modes.} However, the former option also seems to
imply a time crystal, while the latter appears to be thermodynamically disfavored if you
assume the final state is stationary. As is clear from \cite{KimKun23}, the
transfer of energy into increasingly high $\ell=m$ modes indicates the buildup of an equitorial disk of
increasingly large radial extent. If this is the case, it might be tempting to conjecture a Grey
Galaxy end state where the central black hole is slowly rotating Kerr-AdS$_4$
or some other slowly rotating stationary hairy BH ($\Omega_H<1$). But this end state is thermodynamically disfavored: when
$\Omega_{H}<1$, the entropy can be
made larger if we throw some of the $\ell=m$ matter for sufficiently large
$\ell$ into the black hole, as can be shown from the first law
\cite{KimKun23}. This argument assumes the central
black hole is stationary however, so that we know the standard form of the first
law of black hole thermodynamics is applicable. If the central black hole is something like
a black resonator instead, then we are back to a time
crystal.

Thus, current evidence makes a large-$N$ time crystal seem like a live option. 
We will not settle the endpoint of slowly rotating Kerr-AdS$_4$ here, however.
Instead, we will now elaborate a connection to the PI,
and carry out a search for time crystals in the zero angular momentum regime. 

\subsection{The Penrose Inequality}\label{sec:PI1}
\begin{figure}
\centering
\includegraphics[width=0.30\textwidth]{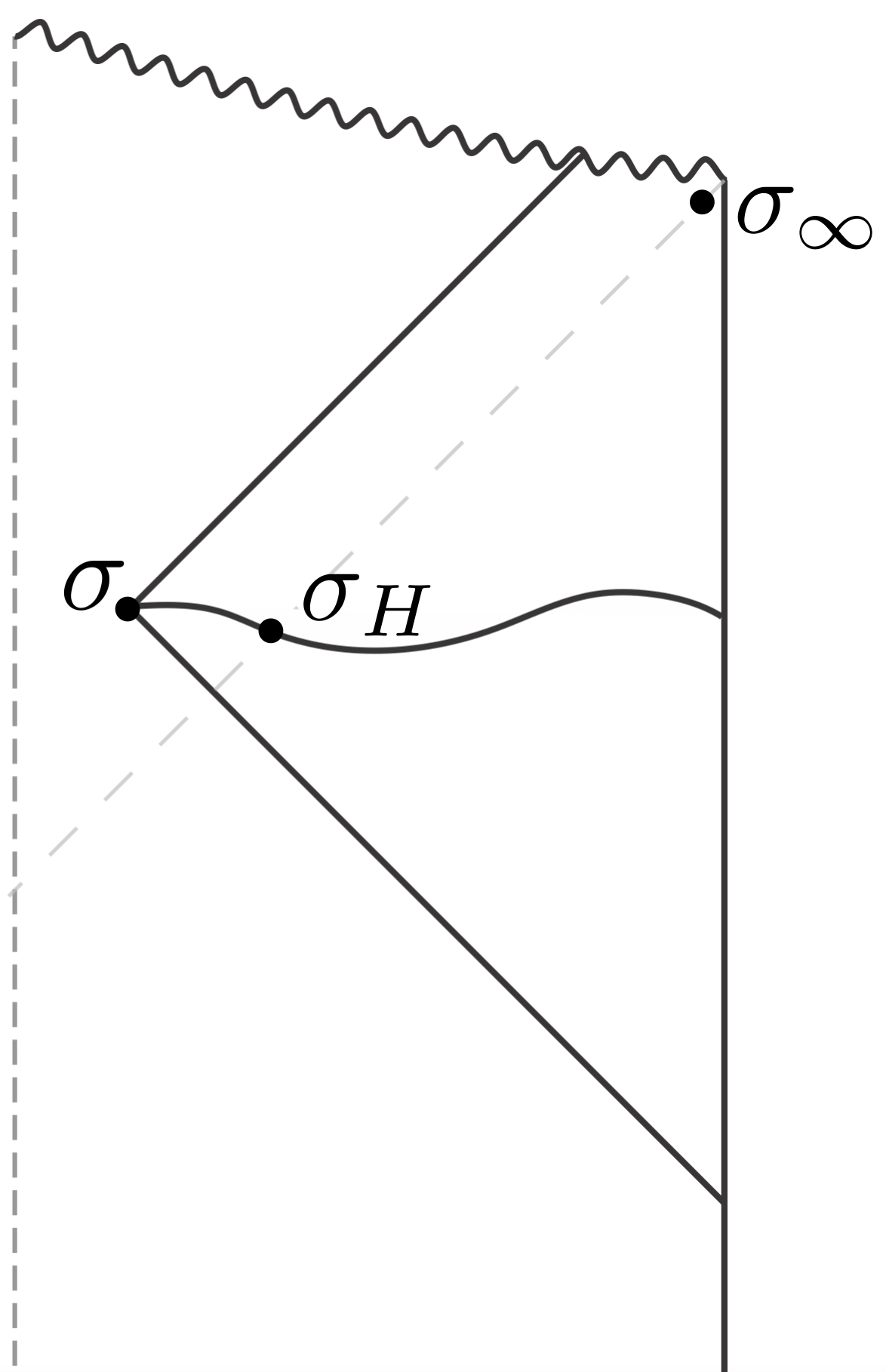}
    \caption{A spacetime of mass $M$, angular momentum $J$, and an outermost
    minimal marginally trapped surface $\sigma$. By outermost minimality $A[\sigma]\leq A[\sigma_{H}]$, and by the
    area theorem $A[\sigma_{H}] \leq A[\sigma_{\infty}]$. By the assumption that
    the black hole settles down to a stationary state   (rather than oscillating
    forever or becoming nakedly singular), $A[\sigma_{\infty}]$
    is the area of a stationary black hole, which in turn is less than the area
    of the most entropic black hole of the same charges.
    }
\label{fig:PIargument}
\end{figure}

The PI is an inequality that was derived by Penrose \cite{Pen73} as a way to test
the weak cosmic censorship conjecture (WCCC) \cite{Pen65}, which states generic gravitational collapse does
not result in naked singularities. Let us review Penrose's original argument \cite{Pen73},
generalizing slightly by allowing for both asymptotically flat (AF) and
asymptotically AdS asymptotics, and the potential existence of hairy black holes.
The quantities involved in the argument below are shown in
Fig.~\ref{fig:PIargument}.

Let $\sigma$ be a compact spacelike codimension-$2$ surface that is marginally
trapped, existing in a spacetime with mass $M$ and angular momentum $J$.
Assuming the null energy condition, we have by the Penrose singularity theorem
\cite{Pen65} that a singularity exists to the future.  Furthermore, assuming the
WCCC, since $\sigma$ is marginally trapped it can be proven that $\sigma$ lies
behind a future event horizon \cite{HawEll,Wald}. Assume now furthermore that
$\sigma$ is homologous to the event horizon and that $\sigma$ is outermost
minimal. The latter means that there exists a spacelike hypersurface $\Sigma$
bounded by $\sigma$ and conformal infinity such that every other codimension-$2$
surface homologous to $\sigma$ in $\Sigma$ has larger area. Let now $\sigma_H$
be the intersection of $\Sigma$ with the event horizon. By outermost minimality
$A[\sigma_{H}] \geq A[\sigma]$. Consider now a cut of the event horizon
$\sigma_{\infty}$ at late times, to the future of $\sigma_{H}$. By the area
theorem \cite{Haw71}, every cut of the event horizon to the future of
$\sigma_{H}$ has larger area.  Assuming the spacetime settles down to a
stationary black hole, $A[\sigma_{\infty}]$ is just the horizon area of a
stationary black hole of mass and angular momentum $M, J$. This area is
trivially smaller than the area of the most entropic black hole with the same
charges, which we denote $A_{\rm stationary}(M, J)$.  Thus, we have
$A[\sigma_{H}] \leq A_{\rm stationary}(M, J)$, finally yielding
\begin{equation}\label{eq:PI}
\begin{aligned}
    A[\sigma] \leq A_{\rm stationary}(M, J).
\end{aligned}
\end{equation}
This inequality is the Penrose inequality -- in 4D AF spacetimes it is often
written as a lower bound on the spacetime mass, which we can do if we assume
that Kerr is the most entropic black hole.  But the form \eqref{eq:PI} is the
most general version. 

Because we had to assume WCCC to derive \eqref{eq:PI}, evidence for the PI is
often considered evidence for WCCC. However, we saw that we also had to assume
that the spacetime settled down to a stationary black hole. Thus, if
\eqref{eq:PI} is found to be false, this could also be considered
evidence in favor of the existence of non-stationary black holes with larger
entropy than all stationary black holes at the same $(M, J)$. In other
words, evidence in favor of time crystals.

What is the status of the PI? No general GR proof exists, but for a moment of
time-symmetry in AF spacetimes, it has been given a proof in \cite{Ger73, HuiIlm01,Bra99,Bra07} assuming the dominant
energy condition. In asymptotically AdS spacetimes, no general proof exists even with
time-symmetry (see \cite{LevFre12,GeGuo13,HusViq17} for more restricted proofs).

In \cite{EngHor19}, Engelhardt and Horowitz took an alternative approach
to deriving the PI.  They were able to derive the PI assuming the AdS/CFT
dictionary, but remarkably not assuming WCCC -- provided you assume a technical
condition on $\sigma$ called stability (see \cite{EngHor19} for details). They
essentially showed that the PI is the bulk manifestation of the following
theorem from statistical mechanics \cite{Schw06}: the microcanonical ensemble on
some energy window is the entropy-maximizing state in the class of states with
support on that energy window. While they got rid of the WCCC assumption, their
derivation did rely on the assumption that the CFT microcanonical ensemble is
dual to a stationary black hole. But this is essentially the assumption that
there are no time crystals.  So the existence of time crystals is a hypothetical
failure mode of all known ways of deriving the PI, whether or not you assume
AdS/CFT. In light of the reviewed recent results on the non-linear instability of slowly
rotating Kerr-AdS$_4$ \cite{FigRos23} and eternally oscillating black holes
\cite{DiaHor11,DiaSan15,IshMur18,IshMur21,Ish21}, it is worth entertaining this
possibility.

Let us remark on a possible confusion at this point. How could the
microcanonical ensemble possibly depend on time, when the state 
\begin{equation}
\begin{aligned}
    \rho_{\rm MC} = \frac{ 1 }{ e^{S} }\sum_{E\in [E_0 - \delta E, E_0 + \delta E]}\ket{E}\bra{E}
\end{aligned}
\end{equation}
is manifestly time-independent? Analogously, from the bulk perspective: if we
found a time-dependent black hole saddle $g$ that dominated the microcanonical
ensemble, then a time-translation of $g$ by a boundary time $T$ would provide a
physically distinct new saddle, since this is a large gauge transformation. And
so in a microcanonical path integral \cite{Mar18}, we would get a one-parameter
family of degenerate saddles we would have to integrate over,\footnote{We thank
Don Marolf for having pointed this out to us, and further pointing out the
possibility for SSB of time translation symmetry, inspiring this work.} restoring time
translation symmetry. Of course, a completely analogous complaint applies for
SSB of spatial translation symmetry. In this case, the
SSB is revealed using a symmetry-breaking field that is turned off
only after the thermodynamic limit is taken. We could try something similar
here. For example, defining $U_{h}(t) = \text{T} e^{-N^2 \lambda \int_{0}^{t}\dd t'
h(t') }$ for some coupling $\lambda$ and time-dependent driving field $h(t)$
independent of $N$,\footnote{We assumed without loss of generality that $1/G_N
\sim N^2$. Replace $N^2$ with $G_N^{-1}$ if you wish.} we could consider the
state $U_{h}(t)\rho_{\rm MC}U_{h}(t)^{\dag}$ and compute observables in this
state, then take the large$-N$ limit and only afterwards send $\lambda \rightarrow 0$.
This likely picks out a particular time-dependent saddle that depends on $h$.
Alternatively, we could avoid a symmetry-breaking field and make the definition
in terms of correlators -- see \cite{WatOsh14} for a proposal. The exact best
way to define a time crystal will not matter to us, so we will not delve into it
further, assuming it can be done.

\subsection{The spherically symmetric Penrose Inequality}\label{sec:PI2}
In the rest of the paper, we will study the spherically symmetric PI in AdS$_4$.
The motivation is the question of whether there exist spherically symmetric time
crystals (or violations of the WCCC). Since the one-point function of the
CFT$_3$ stress tensor must be constant in both time and space with spherical
symmetry,\footnote{In the canonical boundary conformal and Lorentz frames.} all
the non-trivial boundary dynamics in a hypothetical time crystal would
have to be carried by matter fields. In the case of bulk real scalars, which
will be our case of interest, we would have time-dependent and
non-equilibrating one-point functions
$\left<\mathcal{O}(t)\right>$.\footnote{Assuming the scalars do not have compact
support for all time. Then we would only see non-trivial effects in two-point
functions and higher. This case seems highly contrived.}

How plausible is such dynamics? Consider first a different case -- a charged
scalar. In this case it is not hard to imagine a hypothetical scenario where
$\left<\mathcal{O}(t)\right>$ oscillates. Consider a hypothetical overextremal non-singular initial dataset. In this case, rather
than forming a naked singularity upon time evolution, it is equally plausible
that the following happens: the scalar condensate gets repulsed from the black
hole when it gets too close, bouncing off it. Then, as
it travels outwards, it in turn reflects off the AdS boundary in finite time and falls back in
again. And so it repeats ad infinitum, leading to an oscillating
$\left<\mathcal{O}(t)\right>$. We might doubt that overextremal initial
data exists, but it is not ruled out. In fact, the existence of overextremal
initial data is closely tied to the charged version of the PI, which just says
that $A[\sigma]\leq A[M,J,Q]$ for an outermost minimal marginally trapped
surface $\sigma$ in a spacetime with charges $(M, J, Q)$. Overextremality
is one way to violate the charged PI, but it is PI violation rather than
overextremality that is the
essential thing here, since PI violation is the more general condition forbidding 
equilibration to a stationary black hole. And in fact, in AF spacetimes the naive charged
PI,
\begin{equation}
\begin{aligned}
A[\sigma] \leq A_{\rm AdS-Reissner-Nordstrom}(M, Q),
\end{aligned}
\end{equation}
is false \cite{KhuWei14}. And with charged perfect fluid matter, unpublished
numerics by the author have produced AF initial datasets
with $Q > M$. That said, we have not checked whether this theory has
 hairy black holes with $Q> M$, so it is not
clear that the proper PI is violated.\footnote{In
AF spacetimes, since there is no reflection off infinity, violation of the
charged PI would imply neither time crystals nor cosmic censorship
violation. The excess charged matter could just travel outwards to timelike or null
infinity forever.}

Let us now return to the neutral case. We see that some kind of repulsive bulk
dynamics ought to be present to prevent the bulk scalar condensate from
falling into the black hole. For minimally coupled Einstein-scalar theory, the only way to achieve this is through the scalar
potential $V(\phi)$. We want $V(\phi)$ to have regions with $V(\phi)<-3$, so that
negative energy densities (beyond the cosmological constant) create effective repulsion.\footnote{While this might appear
to be unphysical for non-practitioners of AdS/CFT, it is not. Negative and even lower unbounded scalar potentials are
very common in top-down constructions \cite{HerHor03}. They need not destabilize
AdS.} As an example, consider a free bulk scalar with mass squared $\mu^2$. In
Schwarzschild-AdS$_{4}$ (SAdS), the effective potential $\mathcal{V}(r)$ for a scalar mode of fixed
angular momentum behaves, at large $r$, as 
\begin{equation}
\begin{aligned}
    \mathcal{V}(r) \sim \left(\mu^2 + 2\right)r^2.
\end{aligned}
\end{equation}
Thus, we see that $\mathcal{V}\sim -r^2$ if $\mu^2 < -2$, giving an
effective repulsive force. Now, for our CFT to have a lower Hamiltonian, we
must have that the so-called Breitenlohner-Freedman bound holds $\mu^2 \geq
\mu_{\rm BF}^2 = -9/4$ \cite{BreFre82}, but the regime $\mu^2 \in (-9/4,
-2)$ is available. This regime corresponds to boundary operators with scaling dimensions $\Delta \in
(1, 2)$. 

More generally, relevant boundary operators, $\Delta < 3$, will be the focus in
this work.\footnote{By the unitarity bound, we always have $\Delta \geq 1/2$.}
Such bulk scalar fields always have $\mu^2 < 0$ and imply violation of the
so-called dominant energy condition (DEC), which states that $T_{ab}u^{a}v^{b}\geq 0$ for
all pairs of future timelike $v^a, u^a$.\footnote{To be precise, when we discuss
the DEC in AdS, we should first allow a cosmological constant to be subtracted
off the scalar potential.} When it does hold, it has
been proven in spherical symmetry that \cite{HusViq17,Fol22}
\begin{equation}\label{eq:NaivePI}
\begin{aligned}
    A[\sigma] \leq A_{\rm Schwarzschild-AdS}(M).
\end{aligned}
\end{equation}
Thus, there can be no time crystals in this case, further explaining why we
focus on DEC-violating theories, of which scalar fields are the most natural
example.

What about hairy black holes? Even if we were to find violations of
\eqref{eq:NaivePI}, which we refer to as the \textit{naive Penrose inequality}
(NPI), it could be that there exist hairy black holes with $A_{\rm hairy}(M)
>A_{\rm Schwarzschild-AdS}(M)$ so that the proper PI \eqref{eq:PI} still holds. For example,
for charged scalars, the work on holographic superconductors \cite{Har08,Har08b,GauSon09} has revealed hairy
black holes with $A_{\rm hairy}(M, Q) \geq A_{\rm Reissner-Nordstrom}(M, Q)$.
Furthermore, for neutral minimally coupled scalars in AdS, the only case where a no-hair theorem is proven
\cite{SudGon02} is when the scalar has
$\mu^2 \geq 0$, which is precisely complementary to our regime of interest.
However, based on numerical evidence, the following no-hair conjecture was given in
\cite{Her06}: no neutral hairy black holes exist in Einstein-scalar theory
provided (1) there exists a positive mass theorem, and (2) the scalar boundary
conditions respect boundary conformal invariance. But in Sec.~\ref{sec:deltaone}, we find a
counterexample to this conjecture in a consistent truncation of M-theory \cite{GauSon09}. To our knowledge, this is
the first such counterexample.\footnote{There are many hairy black
holes, such as \cite{TorMae01,MarTro04,Zlo04,HerMae04}, but as pointed out in \cite{Her06}, these exist in
unstable theories.} While these black holes do not dominate the
microcanonical ensemble, it shows that we have to worry about hairy black holes.
If we find a violation of the NPI in some theory, we have to map out the hairy
black holes of that theory before we can claim a violation of the PI.

We now proceed to our theories of interest.

\section{The Penrose Inequality in Designer Gravity}\label{sec:designergrav}
\subsection{Scalar fields with general boundary conditions}\label{sec:designer1}
We now consider the four-dimensional action
\begin{equation}
\begin{aligned}
    I = \frac{ 1 }{ 8\pi G_N }\int \dd^{4}x \sqrt{-g} \left[\frac{ 1 }{ 2 }R -
    \frac{ 1 }{ 2 }\nabla_a \phi
    \nabla^a \phi -V(\phi) \right].
\end{aligned}
\end{equation}
We assume that our AdS vacuum of interest is at $\phi = 0$. We pick units where
$L=1$ and assume $V$ is analytic near $\phi=0$, so that
\begin{equation}
\begin{aligned}
    V(\phi) = -3 +\frac{ 1 }{ 2 }\mu^2 \phi^2 + g_3 \phi^3 + g_4 \phi^4 +
    g_5\phi^5 + \mathcal{O}\left(\phi^{6}\right).
\end{aligned}
\end{equation}
A near boundary analysis in standard global coordinates gives that
\cite{HenMar06}\footnote{The dots refer to subleading terms irrelevant for
the computation of charges.} 
\begin{equation}\label{eq:phifalloff}
\begin{aligned}
    \phi &= \frac{ \alpha }{ r^{\Delta_-} } + \frac{ b_1 \alpha^2 }{
        r^{2\Delta_{-}} } + \frac{ b_2 \alpha^3 }{
        r^{3\Delta_{-}} } + \frac{ b_3 \alpha^4 }{
        r^{4\Delta_{-}} } + \frac{ \beta }{ r^{\Delta_{+}} } + \ldots, \\
    \Delta_{\pm} &= \frac{ 3 }{ 2 } \pm \sqrt{\frac{ 9 }{ 4 } +
    \mu^2},
\end{aligned}
\end{equation}
where $\alpha$ and $\beta$ depend on coordinates on the conformal boundary.
The coefficients $b_i$ are completely fixed by $\Delta_-, g_3, g_4, g_5$ -- see
Appendix~\ref{sec:coefs} for explicit expressions. 
The $b_i$-terms are of no relevance when $\mu^2 \geq \mu_{\rm BF}^2 +1$ ($\Delta_-
< 1/2$), since in this case only $\alpha=0$ gives normalizable solutions.

We have
also assumed $\Delta_{-} \notin \{\frac{ 3 }{ 5 }, \frac{ 3 }{ 4 }, 1, \frac{ 3
}{ 2 } \}$, since if $\Delta_{-}$ takes one of these values, then logarithms
appear in \eqref{eq:phifalloff} for generic values of the couplings
\cite{HenMar06}, and special treatment is needed. When we cross these special
values of $\Delta_{-}$, the number of $b_{i}$--terms
that dominate over $1/r^{\Delta_+}$ changes, and the
Hamiltonian generator of the Einstein-scalar system picks up additional terms \cite{HenMar06}. As a
consequence, the range $\Delta_{-} \in (\frac{ 1 }{ 2 }, \frac{ 3 }{ 2 })$,
corresponding to $\mu_{\rm BF}^2 < \mu^2 < \mu_{\rm BF}^2 + 1$,
divides into four distinct regimes with different divergence
structures. These four ranges are $(\frac{ 1 }{ 2 }, \frac{ 3 }{ 5 })$, $(\frac{ 3 }{ 5 }, \frac{ 3 }{ 4 })$,
$\left(\frac{ 3 }{ 4 }, 1 \right)$, $(1, \frac{ 3 }{ 2 })$.

To completely fix a theory, we must specify boundary conditions, consisting of a
functional relationship $\beta=\beta(\alpha)$ \cite{HerMae04,HerHor04b}. 
As mentioned, if $\mu^2 \geq \mu_{\rm BF}^2 +1$ only $\alpha=0$ is possible.
However, in the range $\mu_{\rm BF}^2 < \mu^2 < \mu_{\rm BF}^2 +1$, which is
the most interesting for us, normalizability of the scalar modes is compatible
with a general function $\beta(\alpha)$, which we parametrize through the
function $W(\alpha)$ as
\begin{equation}
\begin{aligned}
    W(\alpha) = \int_{0}^{\alpha}\dd\alpha' \beta(\alpha').
\end{aligned}
\end{equation}
To see the interpretation of $W$ on the CFT side, let $I_{\rm CFT}$ be the CFT action for the
theory with $\beta=0$ ($W=0$), where
$\phi$ is dual to a scalar primary $\mathcal{O}$ of dimension $\Delta_-$. In
\cite{Wit01b} it was argued that boundary conditions characterized by $W$ on the
boundary correspond to a deformation 
\begin{equation}
\begin{aligned}
    I_{\rm CFT} \rightarrow I_{\rm CFT} - \int d^3 x\sqrt{-g} W(\mathcal{O}).
\end{aligned}
\end{equation}
Theories with general $W$ are known as ``designer gravity''
\cite{HerHor04b,HerHol05,AmsMar06,AmsHer07,HerMae04}, since one can 
always design $W$ so as to produce a theory with a ground state soliton of the desired
energy \cite{HerHor04b}.

We will work with the most general
boundary conditions that still preserve conformal symmetry on the boundary, i.e.
which preserve an $\text{SO}(2,2)$ asymptotic symmetry. This requires that  \cite{HerMae04, HenMar06}
\begin{equation}\label{eq:betaalphaabs}
\begin{aligned}
|\beta| = |f\alpha|^{\frac{ \Delta_{+} }{ \Delta_{-} } }
\end{aligned}
\end{equation}
where $f$ is any real constant. Thus, we need either $\beta = f
|\alpha|^{\frac{\Delta_{+} }{ \Delta_{-} }}$ or  $\beta = f \sign(\alpha)
|\alpha|^{\frac{\Delta_{+} }{ \Delta_{-} }}$. 
The former choice results in lower unbounded
energy, and so we work with boundary condition
\begin{equation}\label{eq:confbc}
\begin{aligned}
\beta = f \sign(\alpha) |\alpha|^{\frac{\Delta_{+} }{ \Delta_{-} }},
\end{aligned}
\end{equation}
which gives
\begin{equation}
\begin{aligned}
    W(\alpha) = \frac{ \Delta_{-} }{ 3 }f|\alpha|^{3/\Delta_{-}}.
\end{aligned}
\end{equation}
We see that $W(\mathcal{O})\propto |\mathcal{O}|^{3/\Delta_{-}}$ indeed
corresponds to a marginal deformation. Note that  ``standard quantization'' ($\alpha=0$)
corresponds to $f=\infty$, while ``alternative quantization'' ($\beta=0$)
corresponds to $f=0$.

It is also possible to fix a relationship $\alpha = \alpha(\beta)$,
and define a corresponding $\hat{W}(\beta) = \int_{0}^{\beta} \dd \beta
\alpha(\beta)$ \cite{Ams12}. In the boundary, if $\hat{\mathcal{O}}$ is the $\Delta_{+}$
operator and $\hat{I}_{\rm CFT}$ the CFT action for the $\alpha=0$ theory, this
is interpreted as a deformation of $\hat{I}_{\rm CFT}$ with the term $-\int
\dd^{3}x\sqrt{-g} \hat{W}(\mathcal{O})$.
However, for a boundary condition where there is a one-to-one correspondence between $\alpha$ and $\beta$, which is the case
for \eqref{eq:confbc}, then there is no distinction between the two approaches from
the bulk perspective, so we focus on the former without loss of generality.

\subsection{Positive mass theorems in designer gravity}\label{sec:designer2}
Next, we need to understand the ground state of our theories. When
does a ground state exist, and when is it given by global AdS$_4$?  In other words, is the
mass bounded below, and is it non-negative? Ultimately we are only interested in
violations of the PI in theories with lower-bounded energy.

It was shown in \cite{HerHor04b} (see also \cite{HerMae04}) that the charge associated to the generator of asymptotic
time translations, i.e. the CFT energy $E$, evaluates to 
\begin{equation}
\begin{aligned}
    8\pi G_N E = \int_{S^2}\dd \Omega \left[ M_{0} + \Delta_{-}\alpha \beta +
    (\Delta_+ - \Delta_-)W\right]
\end{aligned}
\end{equation}
where the integral is over a unit sphere at infinity. $M_0$ is the finite
part of the standard uncorrected gravitational Hamiltonian density, which for
$\alpha\neq 0$ has divergent pieces. In spherical symmetry, $M_0$ is the coefficient of the
$\mathcal{O}(1/r^{5})$ term in $g_{rr}$.\footnote{See
\cite{HerMae04} for the exact coordinates. For AdS-Schwarzschild, we have
$g_{rr} =(1+r^2-M_0/r)^{-1}$.} In later sections, where we work in spherical
symmetry, we frequently refer to $M
\equiv 8\pi G_N E / (4\pi)$ as ``the mass''. For
conformal boundary conditions and spherical symmetry, a simple calculation
yields
\begin{equation}
\begin{aligned}
M = M_0 - \frac{ 2 }{ 3 }\mu^2 f |\alpha|^{3/\Delta_-}.
\end{aligned}
\end{equation}

Is $E$ bounded below? 
In case of standard quantization, Witten-Nester type spinorial techniques \cite{Wit81,Nes81} can be used to show
\cite{Bou84,Tow84} that a sufficient condition for a positive mass
theorem (PMT) is that there exist a function $P(\phi)$ such that
\begin{equation}\label{eq:VWpot}
\begin{aligned}
    V(\phi) = 2P'(\phi)^2 - 3P(\phi)^2.
\end{aligned}
\end{equation}
It is also necessary that $P(\phi)$ is defined for all $\phi \in \mathbb{R}$ and satisfies
$P'(0)=0$. It has not been proven that the
existence of $P$, which we refer to as the superpotential,\footnote{We do not
assume supersymmetry, although if we have supersymmetry, $P$ would be the usual
superpotential.} is a necessary
condition, although it has been suggested to be true \cite{Her06,AmsHer07}.
We will find further evidence supporting this, contrary to previous speculation
by the author in \cite{Fol22}. 

For $\alpha\neq 0$ boundary conditions, the situation is more interesting. Let us
without loss of generality fix the sign $P(0)=1$. Then, solving
\eqref{eq:VWpot} for $P$, one finds that there are two classes of solutions
\cite{Pap06,Pap07}. One solution is analytic near $\phi=0$:
\begin{equation}
\begin{aligned}
    P_{+}(\phi) = 1 + \frac{ \Delta_{+} }{ 4 }\phi^2 + \mathcal{O}(\phi^3).
\end{aligned}
\end{equation}
The second class of solutions is a one-parameter family of solutions that is
non-analytic near $\phi=0$, giving\footnote{There is actually a third branch,
discussed in the appendix,  where $|\phi|^{3/\Delta_-}\rightarrow \sign \phi|\phi|^{3/\Delta_-}$. Also, note that $s\neq0$ is only possible
for
$\mu_{\rm BF}^2 \leq \mu^2 < \mu_{\rm BF}^2 + 1$.} 
\begin{equation}
\begin{aligned}
    P_{-}(\phi; s) = 1 + \frac{ \Delta_{-} }{ 4 }\phi^2  - s \frac{
        \Delta_{-}(3-2\Delta_{-}) }{ 6 }|\phi|^{\frac{ 3 }{ \Delta_{-} }} +
    \ldots,
\end{aligned}
\end{equation}
where we have displayed only the leading non-constant analytic and
non-analytic terms.  A formal series solution exists for any value of $s$ and for generic couplings.
However, numerically integrating \eqref{eq:VWpot} to finite values of $\phi$, one finds that $P_{-}(\phi;s)$ only
exists for all $\phi$ when $s$ lies below a critical value $s_c \in \mathbb{R}$ that depends on
the full form of $V(\phi)$. For some potentials no solution exists for any $s$.

It was realized in \cite{AmsHer07} that the existence of $P_+$ does not guarantee
a bound on the mass when $\alpha\neq 0$. Instead, building on \cite{HerHol05,
AmsMar06}, they showed that if $P_{-}$ exists for $s=0$, then
\begin{equation}\label{eq:Wbnd0}
\begin{aligned}
    8\pi G_N E \geq (\Delta_+-\Delta_-)\int_{S^2}\dd \Omega W(\alpha),
\end{aligned}
\end{equation}
provided that the cubic and quintic couplings vanish ($g_3=g_5=0$) whenever $\Delta_- \leq
1$. In the case of conformal boundary conditions, where $W\sim
f|\phi|^{3/\Delta_{-}}$, $f\geq 0$ is a sufficient condition for $E\geq0$
as long as $s_c \geq 0$. 

There are cases where the mass is non-negative even when $f<0$.
It was proven in \cite{FauHor10} (see also \cite{Ams12}) that if $P_-(\phi;s)$ exists for
$s\leq s_c$, and if $V$ satisfies conditions we describe below, then
\begin{equation}\label{eq:Wbnd}
\begin{aligned}
    8\pi G_N E \geq (\Delta_+-\Delta_-)\int_{S^2}\dd \Omega \left[W(\alpha) + \frac{
        s_{c}\Delta_{-} }{ 3 }|\alpha|^{3/\Delta_{-}}
    \right].
\end{aligned}
\end{equation}
Conformal
boundary conditions then give 
\begin{equation}\label{eq:Wbndconf}
\begin{aligned}
    8\pi G_N E \geq \frac{ \Delta_-(\Delta_+ - \Delta_-) }{ 3 } \int_{S^2} \dd
    \Omega(f + s_{c} )|\alpha|^{3/\Delta_{-}},
\end{aligned}
\end{equation}
so in the regime of applicability, $f\geq -s_{c}$ guarantees $E\geq 0$. 

As for the conditions on $V$, the proof presented in \cite{FauHor10} holds
generally when $\Delta_{-} \geq 1$. If the cubic coupling vanishes, then it
also holds for  $\Delta_- \in \left(3/4, 1\right)$.\footnote{If these do not
hold, then (A.11) in \cite{FauHor10} needs to be modified.} It is plausible that
their result holds more generally with some straightforward modifications
of their proof to account for extra divergences. The same can be said for \eqref{eq:Wbnd}
in the range $\Delta_{-}\in(\frac{ 1 }{ 2 }, 1)$ when cubic or quintic couplings
are present. However, there is one interesting fact to be aware of.
When $\Delta_{-} = 1$, we find that no $P_{-}$ superpotential can exist for any
non-zero cubic coupling, since the cubic term of $P_0$ scales as
$g_3/(\Delta_{-}-1)$ (see Appendix ~\ref{sec:superpotapp}). In light of this
and the fact that no PMT is proven for $\Delta_{-} < 1$ when $g_3\neq 0$, it is
interesting to wonder if new effects arise in this regime.

\subsection{ODEs for mass-minimizing initial data}\label{sec:newODE}
Conveniently, we can test PI with just initial data on a spacelike slice $\Sigma$
bounded by $\sigma$ and conformal infinity. The full spacetime is not needed.
Using this fact, we now want to attempt to construct for initial data that violates the PI in spherical symmetry.

There are two natural ways to proceed. Let $r_*$ be the radius of a marginally
trapped surface $\sigma$. To try to violate
\begin{equation}
\begin{aligned}
    4\pi r_*^2 \leq A_{\rm static}(M),
\end{aligned}
\end{equation}
we can either maximize $r_*$ over the space of initial data with
fixed $M$, or we can minimize the mass over the space of initial data with
fixed $r_*$. The latter is true since, by the first law of black hole
thermodynamics, $A$ increases with $M$. We will take the mass-minimizing approach. However, we will
make one reasonable assumption: that mass-minimizing initial data can be
realized by an initial dataset corresponding to a moment of time-symmetry. This is intuitive. To minimize the
mass, we want to add a lump of scalar field that makes the scalar potential
negative over a large spacetime region, while at the same time avoiding paying a
large cost in positive gradient energy. Time-derivatives $\partial_t \phi|_{\Sigma}$ are independent
from $\phi|_{\Sigma}$, and only the latter influences the scalar potential, so
turning on $\partial_t \phi|_{\Sigma}$ adds wasteful kinetic energy. That said, this
reasoning is not rigorous.  When solving the constraint equations in
spherical symmetry, while
direct contributions to the mass from $\partial_t \phi$ terms are positive
definite, contributions from terms with the extrinsic curvature $K_{ab}$ are not all
manifestly positive. However, a more limited result can easily be proven: 
whenever $\Sigma$ is a maximal or minimal volume slice ($K_a^a=0$), then the
contributions from $K_{ab}$ are manifestly positive,\footnote{See for example (3.13) in \cite{EngFol21}} and for
every initial-dataset with $K_{a}^a=0$, there exists a time-symmetric initial dataset with
smaller mass and the same horizon area. Thus, if we miss out on hypothetical time crystal initial data due
to this assumption, these spacetimes must have the following property:
there exist no maximal volume slice anchored at $\sigma$ and the conformal
boundary. This would be very surprising. In the absence of a naked singularity
terminating the spacetime outside the black hole, which we assume when we talk about
a time crystal, this would suggest an infinite volume growth to
the future à la de Sitter.  This seems unlikely in an AdS spacetime formed by
reasonable matter. Note however that if our opinion is that WCCC violation in spherical
symmetry is likely, then it would be good to investigate the PI for initial data with
$K^a_a\neq 0$ as well.

Let us now proceed under our assumption, choosing coordinates
\begin{equation}\label{eq:coords}
\begin{aligned}
    \dd s^2|_{\Sigma} = \frac{ \dd r^2 }{ 1 - \frac{ m(r) }{ r } } + r^2 \dd \Omega^2,
    \quad r\in [r_{*}, \infty),
\end{aligned}
\end{equation}
where $m(r)$ is a function determined by the Hamiltonian constraint. 
Here $r_*$ is the location of the apparent horizon while $r=\infty$ is the
conformal boundary. It is easy to show that the 
coordinates \eqref{eq:coords} break down for some $r=\tilde{r}>0$ if and only if
$m(\tilde{r})=\tilde{r}$, which under our time-symmetry
assumption is equivalent to $r=\tilde{r}$ being an extremal surface (and thus
also
marginally trapped). Thus, we have the boundary condition $m(r_*)=r_*$. Note
that we also assume that one coordinate patch covers $\Sigma$--i.e. that
$m(r)>r$ for all $r>r_*$. This is equivalent to the physical condition that $\sigma$ is
outermost minimal, which is what we want.

Under time-symmetry, the full constraint equations reduce to the Hamiltonian
constraint, which becomes
\begin{equation}
\begin{aligned}
    \mathcal{R} = \nabla^a \phi \nabla_a \phi + 2V(\phi),
\end{aligned}
\end{equation}
where $\mathcal{R}$ is the Ricci scalar of \eqref{eq:coords}. In our
coordinates, the constraint
equation reduces to \cite{HerHor04c}
\begin{equation}
\begin{aligned}
    m' + \frac{ r }{ 2 }m(\phi')^2 = r^2\left[V(\phi) + \frac{ 1 }{ 2
    }(\phi')^2\right],
\end{aligned}
\end{equation}
which is readily integrated to yield the solution
\begin{equation}\label{eq:mr}
\begin{aligned}
    m(r) &= e^{-\frac{ 1 }{ 2 }\int_{r_*}^{r}\dd z z \phi'(z)^2
    }
    \left[r_* + \int_{r_*}^{r}\dd \rho
    e^{\frac{ 1 }{ 2 }\int_{r_*}^{\rho}\dd z z \phi'(z)^2}\rho^2
    \left(V(\phi(\rho))+ \frac{ 1 }{ 2 }\phi'(\rho)^2\right) \right].
\end{aligned}
\end{equation}
If the scalar field has compact support for $r\leq R$, then the
mass of the spacetime is \cite{HerHor04c}
\begin{equation}\label{eq:MR}
\begin{aligned}
   M = m(R) + R^3.
\end{aligned}
\end{equation}
The last term subtracts a diverging $-r^3$ term in $m(r)$ due to the
cosmological constant.
If we work with standard quantization ($\alpha=0$), then the large$-R$ limit of the above
gives the mass when the scalar has non-compact support. With more general
boundary conditions, there are additional scalar-dependent divergent terms in
$m(r)$, and the conserved quantity corresponding to the generator of
asymptotic time translations gets additional contributions from the scalar
field. In particular, at large $r$ we have the behavior \cite{HenMar06}
\begin{equation}\label{eq:mdiv}
\begin{aligned}
    m(r) = M_0 + r^3\left[c_1 \alpha^2r^{-2\Delta_{-}}+c_2 \alpha^3
    r^{-3\Delta_-}+c_3 \alpha^4r^{-4\Delta_-}+c_4 \alpha^{5}r^{-5\Delta}\right] + \ldots
\end{aligned}
\end{equation}
where dots indicate terms decaying as $r \rightarrow \infty$. The coefficients
$c_i$ were computed in \cite{HenMar06}, and we reproduce them in Appendix~\ref{sec:coefs}.

Remember now that the mass is 
\begin{equation}\label{eq:Mcorr}
\begin{aligned}
M = M_0 - \frac{ 2 }{ 3 }\mu^2 f |\alpha|^{3/\Delta_{-}},
\end{aligned}
\end{equation}
and view $M$ as a non-local functional of $\phi(r)$ at fixed $r_*$. We are
looking for stationary points of $M$ with respect to variations of $\phi$. This
includes stationarity with respect to compactly
supported variations. Since \eqref{eq:Mcorr} differs
from $m(\infty)$ only by boundary terms involving the scalars, we get the same equations of motion
by varying \eqref{eq:mr} with respect to $\phi$ as if we varied
\eqref{eq:Mcorr}.

The variation of $m(\infty)$ with respect to $\phi$ produces an integro-differential equation that mass-minimizing
initial data must satisfy. The variation is straightforward though algebraically involved. Introducing the shorthands
\begin{equation}
\begin{aligned}
    \Gamma(r) &= e^{\frac{ 1 }{ 2 }\int_{r_*}^{r}\dd z z \phi'(z)^2 }\\
    H(r) &= \int_{r_*}^{r}\dd \rho \rho^2 \Gamma(\rho) \left[V(\phi(\rho)) + \frac{ 1
    }{ 2 }\phi'(\rho)^2 \right]
\end{aligned}
\end{equation}
we get, after some interchanged integrals and variable renamings, the following
integro-differential equation:\footnote{ A similar equation was derived in
\cite{HerHor04c}, although dropping the scalar gradient term in $m(\infty)$
and with $r_* = 0$. In the setup considered in \cite{HerHor04c}, they wanted to
look for $M<0$ data sets, and a scaling argument let them get rid of the gradient
term. This is not the case for us, since (a) we have $\alpha \neq 0$ BCs and (b)
we want to look at the PI and not just the PMT. Either of these facts make the
scaling argument inapplicable.  }
\begin{equation}
    \begin{aligned}\label{eq:maineq}
    (\phi' + r \phi'')\left[r_* + H(r)\right] + r^2 \Gamma(r)\left[V' + r\phi'V
    -\frac{ 2\phi' }{ r } - \phi'' \right] = 0.
\end{aligned}
\end{equation}

This equation is readily
converted into an equivalent system of ODEs by treating $H$ and $\Gamma$ 
as their own variables to be solved for, and supplementing \eqref{eq:maineq} with the equations
\begin{equation}\label{eq:maineq2}
\begin{aligned}
    H' &= r^2 \Gamma\left[V+\frac{ 1 }{ 2 }(\phi')^2\right], \\
    \Gamma' &= \frac{ 1 }{ 2 }r (\phi')^2 \Gamma,
\end{aligned}
\end{equation}
and the boundary conditions
\begin{equation}
\begin{aligned}
    H(r_*) &= 0, \\
    \Gamma(r_*) &= 1.
\end{aligned}
\end{equation}

To solve equations \eqref{eq:maineq} and \eqref{eq:maineq2}, we pick a value for $\phi_0 \equiv
\phi(r_*)$ and integrate our system outward to larger $r$. Then one of two
things happens. Either the solution exists for all $r$, and $\phi$ approaches an
extremum of $V$ as $r \rightarrow \infty$, or the solution blows up at finite
$r$. In the former case, if $\lim_{r \rightarrow \infty}\phi(r) = 0$, then
we have a solution of interest. The values of $\alpha$ and $\beta$ for this
solution, and thus
also $f = \beta/\left[\sign\alpha|\alpha|^{\Delta_{+}/\Delta_{-}}\right]$, are not fixed
in advance, but depend on the value of $\phi_0$. So solutions of interest induce a map
\begin{equation}
\begin{aligned}
    \phi_{0} \mapsto (\alpha, \beta).
\end{aligned}
\end{equation}
over a finite range of $\phi_0$ values. As we vary $\phi_0$ we
also vary $f$, and so we produce a one-parameter family of solutions where each
solution generically exists in different theories. 

There is also a second class of interesting solutions. Consider a case where the
scalar field has a root $\phi(R)=0$. This can be converted to a solution of
interest by manually truncating the solution at $r=R$, setting $\phi(r>R)=0$.
This produces a continuous solution that is not differentiable at $r=R$. This is
not an issue. Since the mass is not sensitive to $\phi''$, we can smooth out
this kink with arbitrarily small cost in the energy. These kinked compactly
supported solutions have $\alpha=\beta=0$, and so they are solutions in any
theory irrespective of the value of $W$. Note however that, unless we have
$\phi'(R)=0$, the kinked solutions are stationary points only with respect to
variations with compact support on $[r_*, R]$. As follows from a scaling
argument presented in \cite{HerHor04c}, when these solutions exist, there will
exist solutions with compact support on $R'>R$ with even lower mass, and as
$R'\rightarrow \infty$ we expect that $M\rightarrow -\infty$. Thus, if a
potential $V(\phi)$ allows compactly supported solutions, then we expect that no
boundary conditions exist such that the mass has a lower bound.

We are now ready to solve \eqref{eq:maineq} numerically.
In Appendices \ref{sec:sc} and \ref{sec:numerics}, we present details on
numerics. The brief summary is the following. 

First, since the coefficients of divergent terms are determined by $\alpha$, we
need a precise determination of $\alpha$. Thus we directly solve for the 
$\mathcal{O}(1)$ function  $\hat{\alpha}(r) \equiv r^{\Delta_{-}}\phi(r)$.
Next, we determine
$s_c$ in two independent ways. One involves directly solving the ODE \eqref{eq:VWpot} for the
superpotential numerically. The other one is by numerically constructing
spherically symmetric static solitons and leveraging the fact that in the large $\alpha$ limit, $\beta_{\rm
soliton}(\alpha) \sim -s_c \sign\alpha|\alpha|^{\Delta_{+}/\Delta_{-}}$
\cite{FauHor10}. This lets us extract $s_c$ from fitting the relationship
$\beta_{\rm soliton}(\alpha)$ at large $\alpha$.\footnote{When $V$ is not $\mathbb{Z}_2$-symmetric, the
$\alpha\rightarrow \infty$ and $\alpha\rightarrow -\infty$ limits give
different proportionality constants for $\beta_{\rm soliton} \propto |\alpha|^{3/\Delta_{-}}$. Only one of them corresponds to
$s_c$. This the solution to this puzzle is explained in the appendix.}
Finding agreement between these independent methods, we gain
confidence that have a reliable extraction of both
coefficients $\alpha$ and $\beta$, which we obtain by fitting $\hat{\alpha}(r)$
to the near boundary expansion \eqref{eq:phifalloff}.
Finally, to compute the mass, we numerically extract $m(r)$ from our solutions at large $r$ and use the fact
that once we know $\alpha$, we know the divergent terms, and so we can use
\eqref{eq:mdiv} to subtract off these, yielding $M_0$. From $M_0, \alpha,
\beta$,
we can readily compute $M$ from \eqref{eq:Mcorr}.

\section{Results}\label{sec:results}
\subsection{Standard quantization}
Consider the theory
\begin{equation}\label{eq:th_standard}
\begin{aligned}
    V(\phi) = -3 - \frac{ 9 }{ 16 } \phi^2 + g_3 \phi^3 + g_4 \phi^4,
\end{aligned}
\end{equation}
which has $\mu^2 = \frac{ 1 }{ 2 }\mu_{\rm BF}^2$. The particular value of the
mass has no significance, beyond it being negative and satisfying $\mu^2 >
\mu_{\rm BF}^2 + 1$, so that only $\alpha=0$ boundary conditions are possible.
The scaling dimension of the dual primary is $\Delta_{+} \approx 2.56$. 

\begin{figure}
\centering
\includegraphics[width=0.99\textwidth]{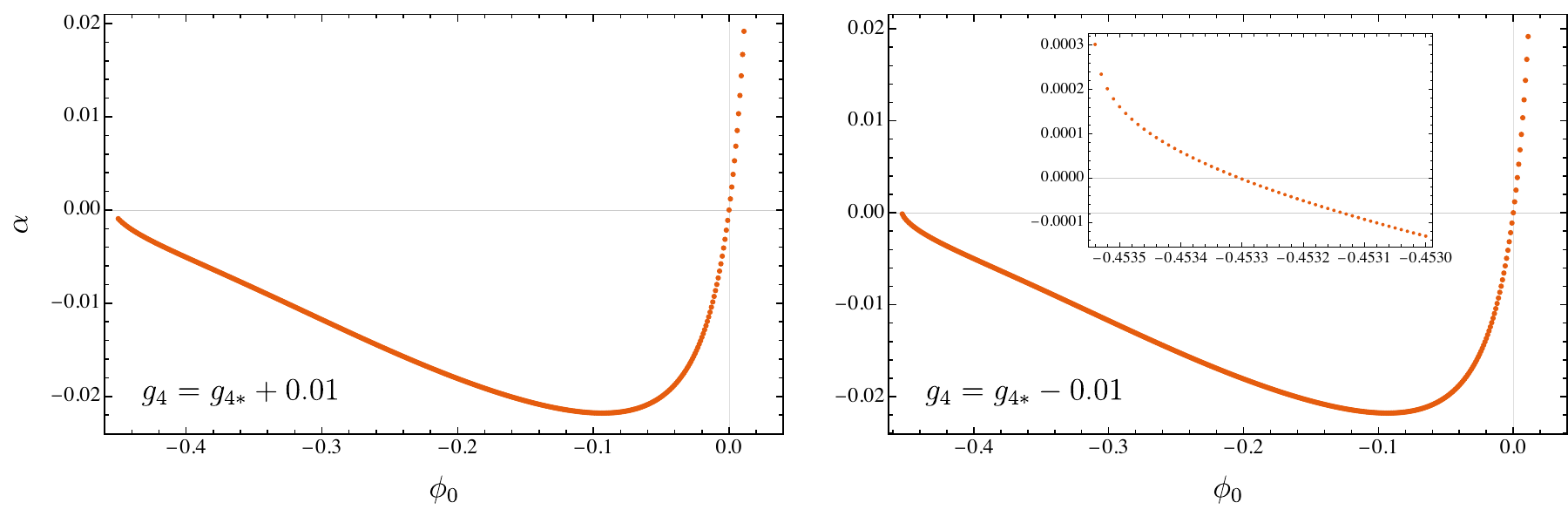}
    \caption{$\alpha(\phi_0)$ for solutions with $r_*=1$ just
    marginally inside (left) and outside (right) the regime
    where a superpotential exists, assuming the potential \eqref{eq:th_standard}
    and $g_3=9$.
    }
\label{fig:standard_plt1}
\end{figure}

\begin{figure}
\centering
\includegraphics[width=0.6\textwidth]{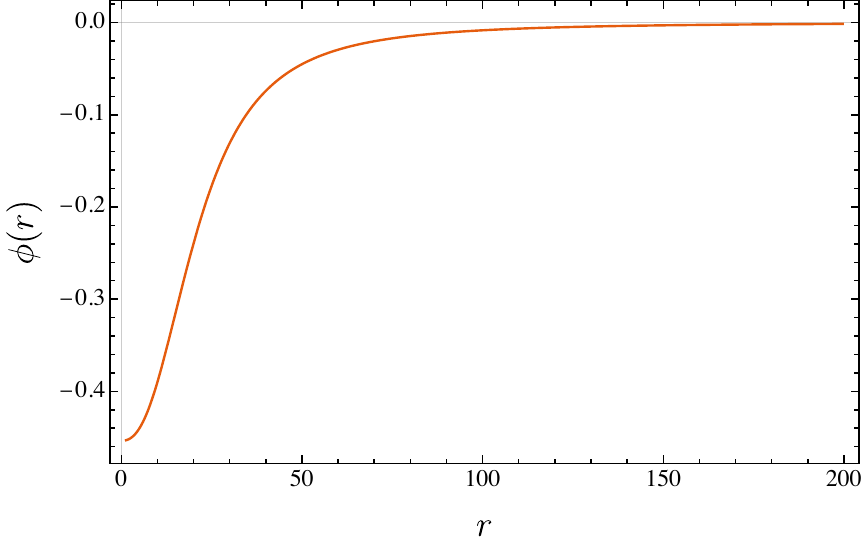}
    \caption{Solution at $r_*=1$ with $\alpha=0$ just marginally inside the regime
    where the superpotential does not exist -- the potential
    \eqref{eq:th_standard} with $(g_3,
    g_4)=(9, g_{4*} - 0.01)$. The initial condition is $\phi_0 =
    -0.45330$, and the total mass is $M = 2.648 > M_{\rm SAdS} = 2$. 
    }
\label{fig:standard_plt2}
\end{figure}
\begin{figure}
\centering
\includegraphics[width=0.99\textwidth]{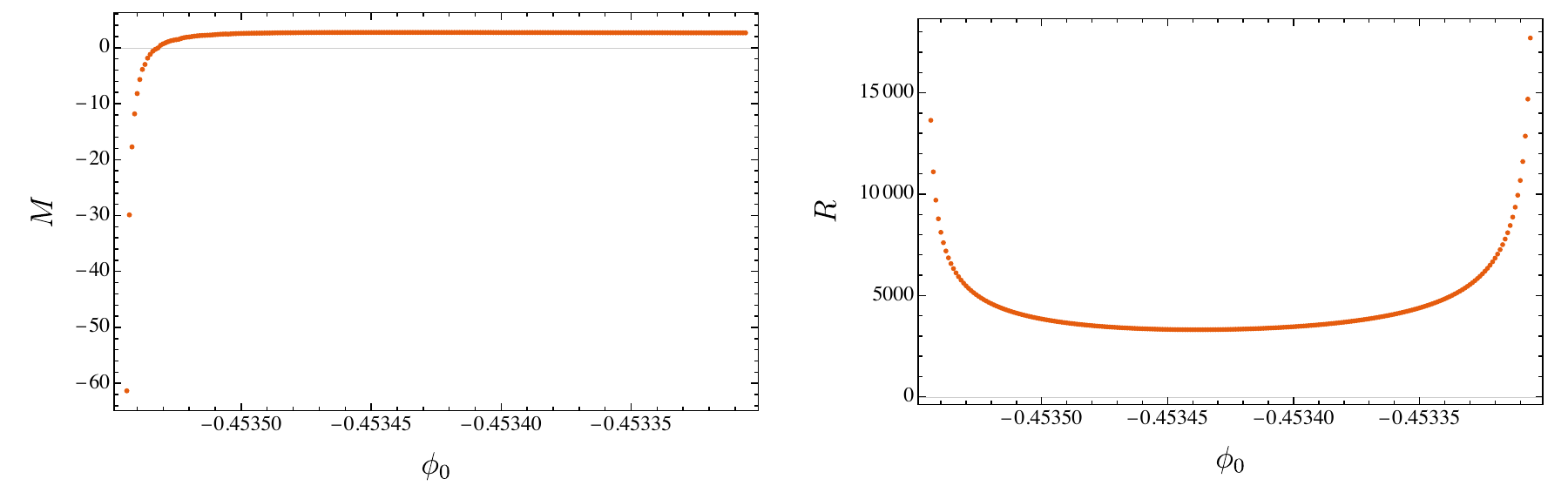}
    \caption{Solutions to \eqref{eq:maineq} with compact support on $[1, R]$ for a
    theory just marginally lacking a superpotential, corresponding to
    \eqref{eq:th_standard} with $(g_3,
    g_4 )=(9, g_{4*} - 0.01)$.
    }
\label{fig:standard_plt3}
\end{figure}

For concreteness and ability to compare with \cite{Fol22}, consider now $g_3=9$. For
this cubic coupling, we have a critical quartic coupling $g_{4*} = 16.26$. A
superpotential exists if and only if $g_4 \geq g_{4*}$. In
Fig.~\ref{fig:standard_plt1}, we plot the relationship $\alpha(\phi_0)$ for
solutions to \eqref{eq:maineq} just marginally inside and outside the regime
where a superpotential exists, at horizon radius $r_*=1$. For values of $\phi_0$ outside the regime
plotted, $\phi$ diverges at finite $r$ or converges to the wrong extremum at
infinity. We see from the left panel of Fig.~\ref{fig:standard_plt1}
that when a superpotential exists, there is 
no non-trivial solution with $\alpha=0$. However, when
there is no superpotential, a non-trivial solution exists -- see the inset in
the right panel of
Fig.~\ref{fig:standard_plt1}. In Fig.~\ref{fig:standard_plt2} we plot this
solution. Computing its mass, we find $M=2.649$, which is greater than the mass
of SAdS with $r_*=1$, which is $M=2$. Thus, this solution respects the
PI. However, when there is no superpotential, we also find a
family of solutions with compact support on the interval
$[1, R(\phi_0)]$ (here are no such solutions for $g_4=g_{4*}+0.01$).
We plot their support $R$ and mass $M$ as a function of $\phi_0$ in
Fig.~\ref{fig:standard_plt3}. We see that these solutions only exists at very
large support: $R\gtrsim 3000$. We also see that as $\phi_0$ approaches $\phi_0 \approx -0.45354$, $R$ blows up while the mass goes
negative and probably diverges to $-\infty$. Thus, while there are violations of
the PI in this theory, they are not interesting since the mass is lower
unbounded. We find similar behavior for other values of $g_3$. 

It thus
appears clear that all the violations of the
PI presented in \cite{Fol22} were for theories with no lower bound on mass. The reason we did not
find these negative mass solutions in \cite{Fol22} is the following: the closer we get to the
regime where $W$ exists, the larger
$R$ needs to be to produce a negative mass solution.\footnote{Or for non-compactly
supported solutions, the more fine-tuned these need to be.} In the case studied
here, thousands of times larger than the scale set by the horizon. Our
numerics in \cite{Fol22} were not sensitive to such types of initial data.

Going forward,
we will not plot $\phi(r)$ again, since the solution always looks qualitatively
like Fig.~\ref{fig:standard_plt2}.

\subsection{$\Delta_ - = 1$}\label{sec:deltaone}
In the regime $\mu^2 < \mu_{\rm BF}^2 +1$, we start by 
considering the special value of $\Delta_{-} = 1$, where (1) we can compare to
existing results, and (2) where we can study some top-down scalar potentials.

We will restrict to $\mathbb{Z}_2$ symmetric potentials here, since it appears
that $g_3 \neq 0$ is incompatible with a lower bounded mass when $\Delta_{-} = 1$
(assuming $\alpha\neq0$). This is because the $P_{-}(\phi; s)$ branch has a
small-$\phi$ expansion \begin{equation} \begin{aligned} P_{-}(\phi; s) = 1 +
\frac{ \Delta_{-} }{ 4 } \phi^2 + \frac{ g_3 }{ 6(\Delta_{-} - 1) }\phi^3 +
\ldots, \end{aligned} \end{equation} and so the $P_{-}$ branch never exists when
$\Delta_{-}=1$ and $g_3\neq 0$. Note also that when $g_3=0$, a logarithmic
branch of the scalar field is absent \cite{HenMar06}, so we can rely on the
lower bounds on mass described earlier.

First we consider
\begin{equation}\label{eq:th1}
    \begin{aligned}
        V(\phi) = - 2 - \cosh(\sqrt{2}\phi).
\end{aligned}
\end{equation}
which is a consistent truncation of a dimensional reduction of
M-theory/11D SUGRA on $S^7$ \cite{CveDuf99}.
In \cite{FauHor10} it was found that this theory has $s_{c} =0$. Our code reproduces $s_c=0$ with both the soliton method and with the
direct method.\footnote{In the special case of $s_c=0$, reproducing $s_c$ with the soliton method just
corresponds to finding that $\beta_{\rm soliton}$ grows slower than
$|\alpha|^{3}$. For \eqref{eq:th1}, $\beta_{\rm soliton}(s)$ just approaches a constant. For the potential
    $V=5/2  - 6 \cosh(\phi/\sqrt{2}) + \cosh(\sqrt{2}\phi)/2$ and $V=-3 -
    \phi^2$, we reproduce the non-zero $s_c$ values from \cite{FauHor10}.} 

\begin{figure}
\centering
\includegraphics[width=0.99\textwidth]{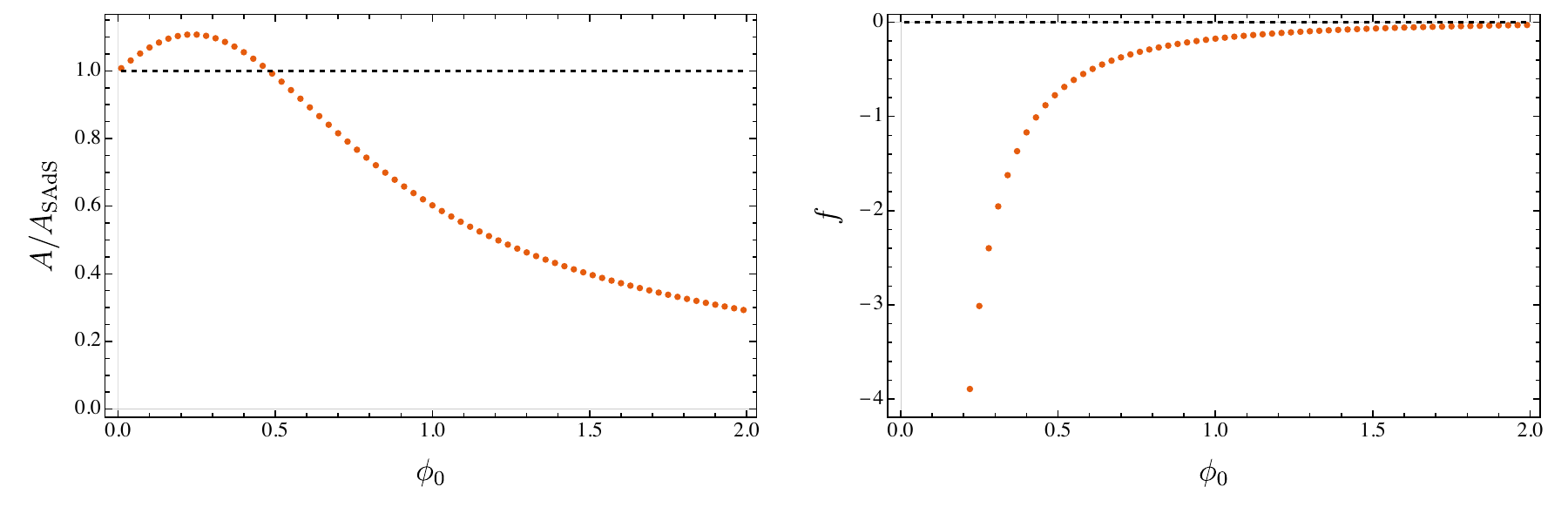}
    \caption{Properties of mass-extremizing solutions with $r_* = 1$ with
    the potential \eqref{eq:th1}. The black dashed line in the right panel shows
    $-s_{*}$, and
    theories with $f<-s_{c}$ have no lower bound on mass.
    For all $\phi_0$ where solutions exist (we do not plot the
    full range), we find $f<-s_{*}$. 
    }
\label{fig:th1_mainplot}
\end{figure}
In Fig.~\ref{fig:th1_mainplot} we display the $f$ value and area ratio of
mass-extremizing solutions as function of $\phi_0$ for $r_* =1$. While there are solutions
that violate the NPI for sufficiently small $\phi_0$, all solutions have $f<-s_c
= 0$, and so the theories in question have no proven lower bound on the energy
and likely lower unbounded mass. The same
qualitative behavior is found for $r_*=0.1$ and $r_* = 10$, and so this theory
seems to respect the PI for all conformally invariant
boundary conditions known to have $M\geq 0$. 

\begin{figure}
\centering
\includegraphics[width=0.99\textwidth]{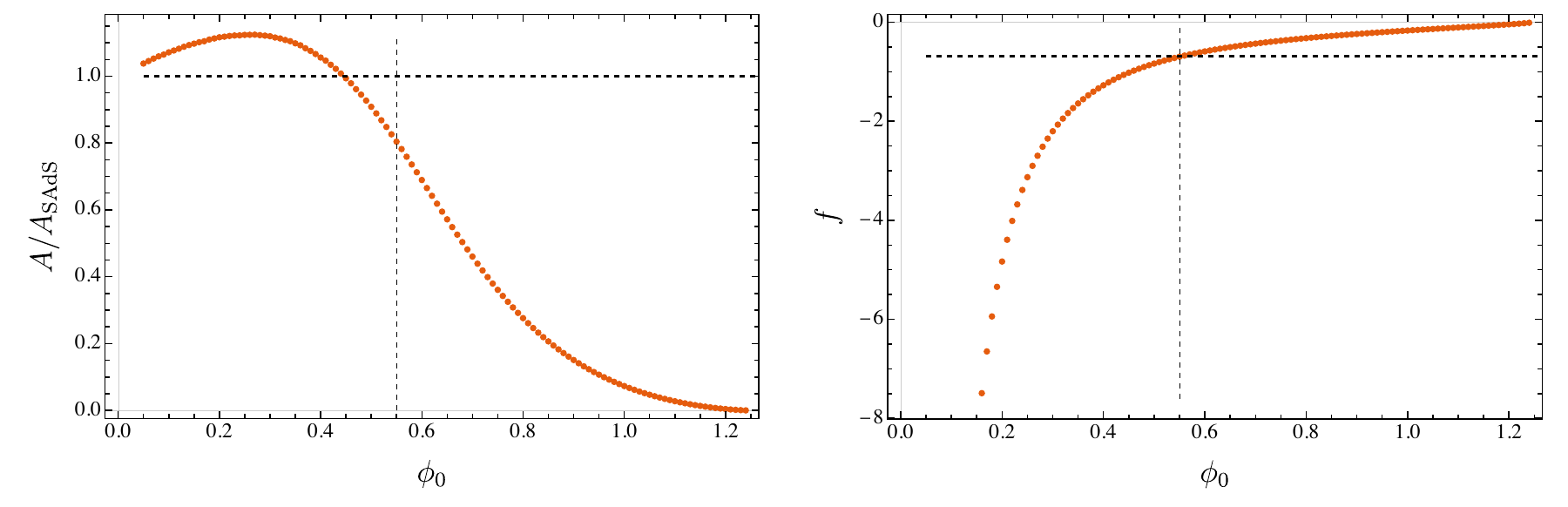}
    \caption{Properties of mass-extremizing solutions with $r_* = 1$ and
    the potential \eqref{eq:th2}. Solutions above the black dashed line in the left
    panel violate the NPI. The black dashed line in the right panel is $-s_c$, and the
    theory in question only has a PMT when $f$ lies above this line. This
    happens at $\phi \approx 0.55$, which we have indicated with a vertical dashed
    line in both plots.
    }
\label{fig:th2_mainplot}
\end{figure}
Next, consider the theory 
\begin{equation}\label{eq:th2}
\begin{aligned}
    V = \frac{ 1 }{ 2 }\cosh\left(\phi/\sqrt{2}\right)^2\left[\cosh(\sqrt{2}\phi) -
    7\right]
\end{aligned}
\end{equation}
which is also a dimensional reduction and consistent truncation of M-theory
\cite{GauSon09}.\footnote{To get this form of the potential, set $\hat{\chi}$ to
be real in Eq.~(7) of \cite{GauSon09} and do the field
redefinition $\hat{\chi} = \sqrt{2} \tanh(\phi/\sqrt{2})$. Then reintroduce the
dimensionalful AdS scale and note that the AdS scale
in their action is $L=1/2$.} We find $s_c = 0.69$. In Fig.~\ref{fig:th2_mainplot} we show properties of
$r_* = 1$ solutions. This time, there are both solutions with a PMT and
solutions that
violate the NPI. However, the regimes have no overlap. $f$
exceeds $-s_c$ only for $\phi_0 \gtrsim 0.55$, but in this range the PI is
respected. The same
qualitative behavior is found for $r_*=0.1$ and $r_* = 10$, and so also this theory
likely respects the PI. 

\begin{figure}
\centering
\includegraphics[width=0.7\textwidth]{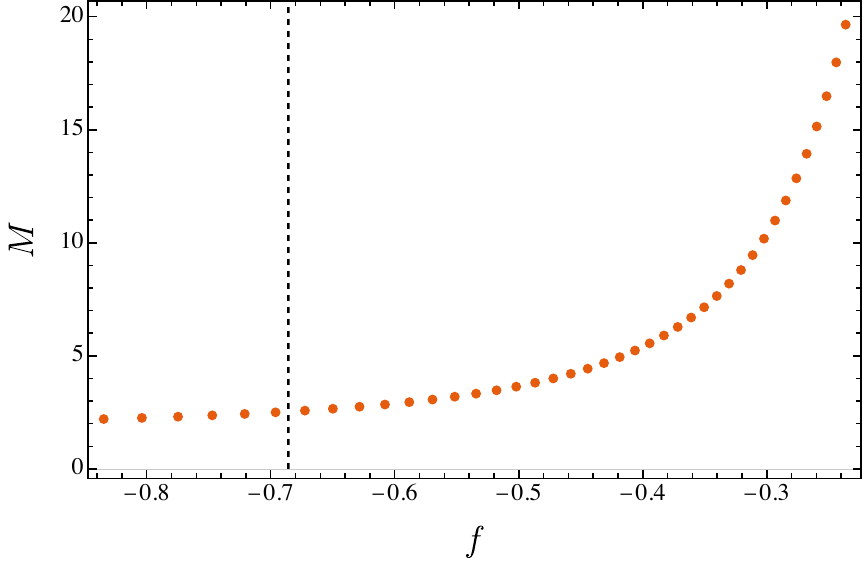}
    \caption{One-parameter family of hairy black holes with $r_*=1$ in the
    theory \eqref{eq:th2}. The vertical dashed line represents $f=-s_c$, and all
    solutions with $f>-s_c$ belong to theories with a PMT. We find that solutions
    exist for $f$ arbitrarily close to $0$ from below, but not for $f>0$. 
    }
\label{fig:th2_bhplot}
\end{figure}
We see from Fig.~\ref{fig:th2_mainplot} that the theory does have non-trivial solutions in the
PMT regime. These cannot dominate the microcanonical ensemble, so a
good guess is that they are subdominant hairy black holes. We now verify this by
constructing hairy black
hole solutions in this theory numerically. The procedure is well known (see for example \cite{Her06}), so we
just present the final result. Fixing $r_*=1$ and varying the scalar on the
horizon, which is analogous to $\phi_0$, we produce a one-parameter family of black
holes with different values for $\alpha, \beta$, which we can use to compute the $f$-value and
mass. In Fig.~\ref{fig:th2_bhplot} we show the resulting $(f, M)$ curve. We
indeed find stable theories with hairy black holes (and they are
microcanonically subdominant since $M>2$). The author is not aware of any
previously constructed neutral spherical hairy black holes in minimally
coupled Einstein-scalar theory that have (1) a
proven PMT, and (2) conformal boundary conditions. The hairy black holes of
\cite{Her06} were in theories without conformal invariance, and 
\cite{Her06} conjectured that neutral hairy black holes do not exist in theories
satisfying (1) and (2). Thus, this appears to be a counterexample to the
conjecture.

We next consider
\begin{equation}\label{eq:th3}
\begin{aligned}
V = -3-\phi^2 + g_4 \phi^4. 
\end{aligned}
\end{equation}
For $g_4 < g_{4*} = -0.74$, no $P_-$ exists for any $s$.  We study
 $g_4 \in \{-0.5, 0, 0.5, 1, 20\}$, at the radii $r_* \in
\{0.1, 1, 10\}$. We find qualitatively similar results
to what we saw in the previous two theories. While there often is both a NPI-violating regime
and a regime where the PMT holds, they do not overlap, even though the
cross-over can be very close. The plots look qualitatively similar to the
first two theories studied in this section, so we do not include them. 

\subsection{$\Delta_- \in (1, 3/2)$}
Now we consider 
\begin{equation}\label{eq:th4b}
\begin{aligned}
    V(\phi) = -3 - \frac{ 35 }{ 32 }\phi^2 + g_3 \phi^3 + g_4 \phi^4,
\end{aligned}
\end{equation}
corresponding to a theory with $\Delta_{-}=5/4$.
We pick this value of the scaling dimension simply because it lies in the center
of the range $\Delta_{-} \in (1, 3/2)$. The $P_-$ superpotential only exists for
the pure cubic theory when
$|g_3| \leq g_{3*}\approx 0.56$, and for the pure quartic theory when
$g_4 > g_{4*} \approx -0.57$.\footnote{Note that since we are so close to the BF
bound, the non-analytic terms in the superpotential series converges very
slowly, as discussed in the appendix. Thus, we are only able to determine one
decimal precisely using this method. The quoted second decimal was obtained using
the soliton method.}

\begin{figure}
\centering
\includegraphics[width=0.99\textwidth]{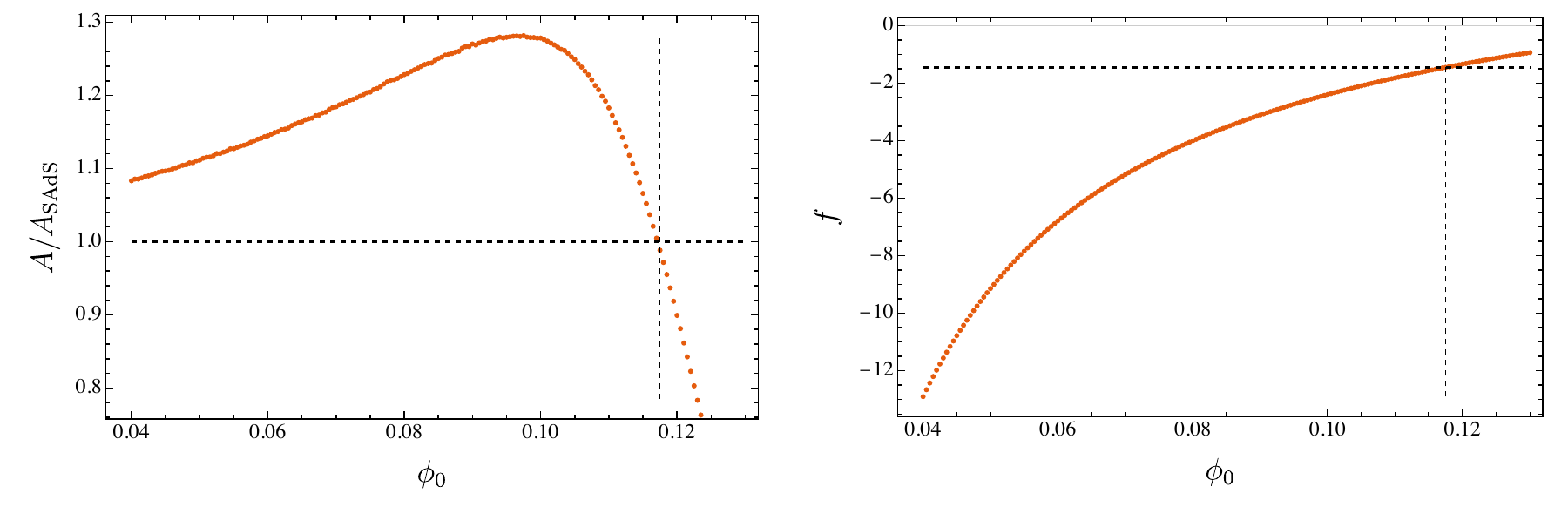}
    \caption{Properties of mass-extremizing solutions with $r_* = 1$ and
    the potential \eqref{eq:th4} with $(g_3, g_4)=(0, 20)$. Solutions above the black dashed line in the left
    panel violate the NPI. The black dashed line in the right panel is $-s_c$, and the
    theory in question only has a PMT when $f$ lies above this line.
    }
\label{fig:th7_mainplot}
\end{figure}
For a pure quartic theory ($g_3=0$) we consider $g_4 \in \{-0.5, 0,
0.5, 1, 20\}$ and $r_* \in \{0.1, 1, 10\}$. We plot the case of
$r_*=1$ and $g_4=20$ in Fig.~\ref{fig:th7_mainplot}, where we find that the regime where the
PMT holds is almost perfectly complenmentary to the regime where the NPI is
violated. Slightly perturbing the parameters $r_*$ and $g_4$ does not
produce violations, since this behavior seems robust. In all other cases we find
results qualitatively similar to the various cases we have already seen, with
the exception of one feature. We find cases where $M<0$. However, these satisfy
$f<-s_c$, as they must. 

We also consider purely cubic theories with $g_3 \in \{0.1, 0.25, 0.5 \}$. We find similar results as earlier -- no PI violations in stable theories.

\subsection{$\Delta_- \in (3/4, 1)$}
Now we consider 
\begin{equation}\label{eq:th4}
\begin{aligned}
    V = -3 - \frac{ 119 }{ 128 }\phi^2 + g_4 \phi^4,
\end{aligned}
\end{equation}
giving a theory with a scaling dimension in the middle of the range
$(\frac{ 3 }{ 4 }, 1)$, meaning $\Delta_- = \frac{25}{64} = 0.875$.
The critical coupling is $g_{4*} = -0.58$. We do not consider a cubic
coupling, since no lower bound on the mass has been proven with a cubic coupling for this range of dimensions.

\begin{figure}
\centering
\includegraphics[width=0.99\textwidth]{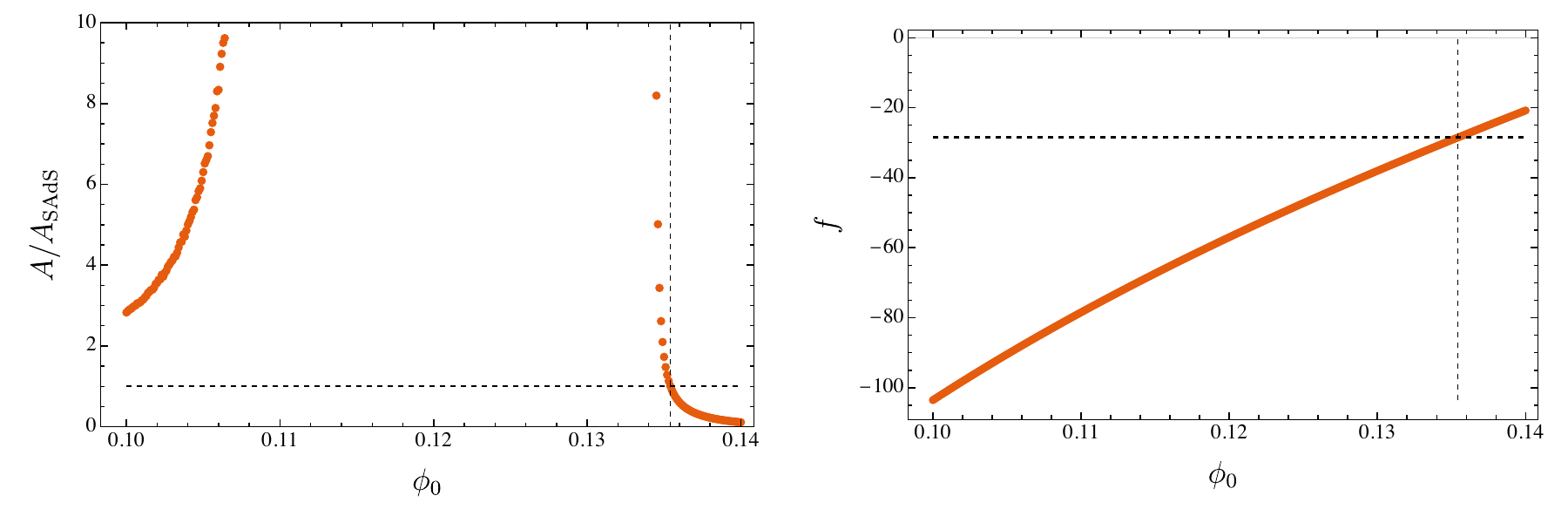}
    \caption{Properties of mass-extremizing solutions with $r_* = 1$ for the
    potential \eqref{eq:th4} with $g_4=20$. Only theories right of the dashed
    vertical line have a PMT. To the left of this line, negative mass solutions
    are present, leading $A/A_{\rm SAdS}$ to be ill-defined.
    }
\label{fig:th4_mainplot}
\end{figure}

We consider $g_{4}\in \{-0.5, 0, 0.5, 1, 20 \}$. We find results qualitatively similar to what we have seen
earlier, except sometimes the mass goes negative, leading $A/A_{\rm
SAdS}$ to diverge.  However, this is always in the $f < -s_c$ regime, and so is
not surprising. See Fig.~\ref{fig:th4_mainplot}.

\subsection{$\Delta_- \in (3/5, 3/4)$}
Now we consider 
\begin{equation}\label{eq:th5}
\begin{aligned}
    V = -3 - \frac{ 2511 }{ 3200 }\phi^2 + g_4 \phi^4
\end{aligned}
\end{equation}
giving a theory with a scaling dimension in the middle of the range
$(\frac{ 3 }{ 5 }, \frac{ 3 }{ 4 })$, meaning $\Delta_- = \frac{27}{40} = 0.675$.
A superpotential only exists for $g_{4}\geq g_{4*} \approx -1$.
We do not consider a cubic or quintic coupling, since no lower 
bound on the mass has been proven with a cubic or quintic coupling for this range of dimensions.
Strictly speaking, even with $g_3=0$, a PMT has only been proven when $f\geq 0$
and $s_c\geq 0$, rather than $f\geq -s_c$. It does however seem quite likely that
the proof of \cite{FauHor10} generalizes, and we will find some numerical
evidence for
this.

\begin{figure}
\centering
\includegraphics[width=0.99\textwidth]{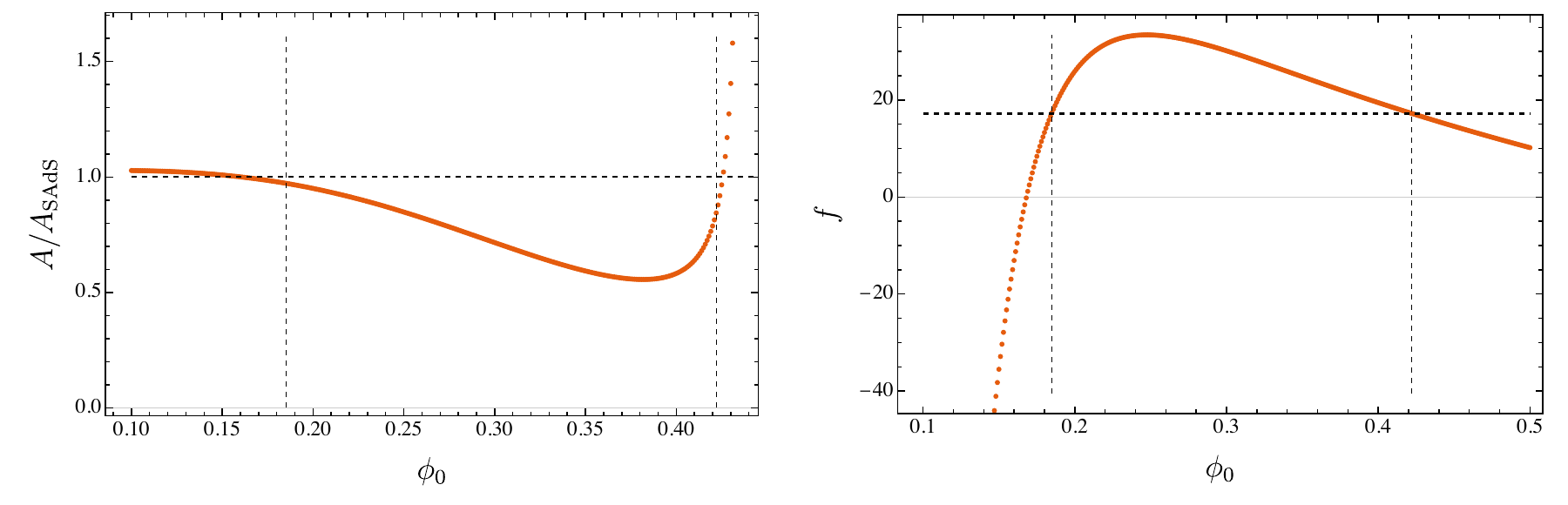}
    \caption{Mass-extremizing solutions for the potential \eqref{eq:th5} with
    $r_* = 1$ and $g_{4}=1$.
    }
\label{fig:th6_mainplot}
\end{figure}

We consider $g_4 \in \{-0.5, 0.2, 0, -0.2, 0.5, 1, 20 \}$, and
find results qualitatively similar to what we have seen earlier. We plot the
case of $r_* = 1$ and $g_4 = 1$ in Fig.~\ref{fig:th6_mainplot}. We see that
negative mass (diverging area ratio) only appears just after we enter the
$f<-s_c$ regime. Similar behavior is found for other $g_4$.

\subsection{$\Delta_- \in (1/2, 3/5)$}\label{sec:th6}
Now we consider 
\begin{equation}\label{eq:th6}
\begin{aligned}
    V = -3 - \frac{ 539 }{ 800 }\phi^2 + \lambda \phi^4
\end{aligned}
\end{equation}
giving a theory with a scaling dimension in the middle of the range
$(\frac{ 1 }{ 2 }, \frac{ 3 }{ 5 })$, meaning $\Delta_- = 11/20 = 0.55$.
A superpotential only exists for $g_4 \geq g_{4*} \approx -1$.
We consider $\lambda \in \{-0.5, -0.2, 0, 0.2, 0.5, 1, 20 \}$. 

We find results qualitatively similar to what we have seen earlier, however note
that we are less confident about our results in this regime that in the previous
regimes. In this regime, some extra care is needed in numerics, since
falloffs are very slow. For example, $M_0$ converges to a constant with a tail
decaying as $\mathcal{O}(1/r^{1/10})$. We find that estimates of $\beta$
stabilize, and thus $s_c$ is reliable, only for $r$ very large, preferably $r\gtrsim 10^5$. 
However, our estimate of $M_0$
starts to get noisy above $r\sim 4\times 10^{4}$, presumably because we manually
are subtracting off divergent terms from $m(r)$ at large $r$ to extract $M_0$, which gets numerically problematic for
very large $r$. We decide to use
$r_{\rm max}= 4\times 10^5$ to estimate $s_c$ via the soliton method, since we
do not need $M_0$ for this computation. However, when computing the mass for our
mass-extremizing solutions we use $r=4\times 10^{4}$. For these values we find
results compatible with the PI. However, a more careful study with improved
numerics in this regime
would be worth doing, since relatively small changes to $s_c$ or $M_0$ in our results would produce a
violation of the PI in the PMT regime.

\section{Discussion}\label{sec:discussion}
In this paper, we have argued that existing candidate solutions \cite{FigRos23} for the
endpoint of the non-linear instability of Kerr-AdS$_4$ are candidates for
being holographic large-$N$ thermal time crystals. Then we pointed out that
holographic time crystals with entropy of order $1/G_N$ imply
violations of the Penrose inequality and carried out a large study of
the AdS$_4$ PI in Einstein gravity coupled to a real scalar field. We focused on scalars
dual to relevant operators in the regime where a large number of different
boundary conditions are possible (``designer gravity''), since this regime was
argued to be most likely to violate the PI. 
Our approach was to derive an ODE system for mass-minimizing initial data at
fixed entropy.\footnote{Assuming mass-minimizing data is time-symmetric. See
main text for arguments why this is reasonable, at least when looking for time
crystals rather than violations of weak cosmic censorship.}
Focusing on boundary conditions compatible with
boundary conformal symmetry, we
found strong evidence that the spherically symmetric PI holds whenever the
Hamiltonian is bounded from below. This suggests that electrically neutral time
crystals in a CFT$_3$ would have to have non-zero angular momentum. 

We also found that earlier violations of the PI by the author in \cite{Fol22}
existed in theories with lower unbounded mass. This unfortunately means that
there are no known examples where the PI serves as swampland condition with
any more constraining power than simply demanding the energy to be lower bounded. 
It could in principle still be true that the PI can function as a Swampland
constraint, albeit the question of time
crystals would have to be settled first.

In Sec.~\ref{sec:deltaone} we showed that there exist neutral hairy black
holes in a consistent truncation of M-theory \cite{GauSon09} with both a
positive mass theorem, and conformally invariant boundary conditions. 
This is, to our knowledge, the first counterexample to the no-hair conjecture of \cite{Her06}.
On the boundary, this theory is a marginal triple trace deformation of the
alternative quantization theory where the scalar is dual to a $\Delta=1$
operator. The hairy black holes and a (provably) lower bounded Hamiltonian only coexist for a finite range of the deformation
parameter ($f \in (-s_c, 0)$). The new BHs do not dominate the microcanonical ensemble, but it
would be interesting to investigate if they dominate the canonical ensemble, or
how these hairy black hole might influence observables away from strict
$N\rightarrow \infty$.

There are interesting paths forward. The most promising, but perhaps also
hardest, is to determine the endpoint of the non-linear instability of slowly
rotating Kerr-AdS$_4$. We could approach this through the Penrose inequality,
rather than with standard time-evolution. Analogous to the approach taken here,
we could try to directly search for initial data that minimizes mass $M$ given a
fixed spin and apparent horizon area. Without spherical symmetry this is a much
harder problem, however, especially since there is no simple explicit functional
that expresses the mass as function of the bulk field profiles. However, if one
has a fast initial data solver and the ability to do efficient deformations of
initial data, perhaps one can use deep learning methods to do gradient descent
on initial data, using the mass as the loss function? 

It would also be interesting to consider charged scalars, since we have
argued that repulsive forces are useful for constructing violations of the PI.
Trying to construct over-extremal spherically symmetric initial data sets in AdS
would be a good place to start. For this one can use similar methods as in this
paper, albeit with the additional complication of a gauge field. This seems
manageable.  It would also be easy to modify this study to work with non-conformal boundary
theories.  It would essentially just require reinterpreting existing solutions with a
modified mass formula.

It would also be worth removing time-symmetry assumption. While this
is unlikely to reveal anything new in the search for time crystals, it might
reveal violations of the PI that are caused by weak cosmic censorship violation. This should
not be too hard, since the constraint equations can still be integrated in this
case. We just get additional equations and terms involving the extrinsic
curvature. 

Finally, unless there exists a novel unknown positive mass theorem that does not
require the existence of a $P_-$ superpotential, we found that a scaling
dimension of $\Delta_{-} = 1$ is incompatible with a non-vanishing cubic
 $g_3$ and a lower bounded Hamiltonian.  It would be interesting to clarify how
 this relates to the story of extremal correlators \cite{DHoFre99}, specifically, the
so-called \textit{shadow-extremal} couplings recently discussed in
\cite{CasMar24}. Also, no positive mass theorems have been proven with
$g_3\neq 0$ and $\frac{ 1 }{ 2 } < \Delta_{-} < 1$. Does anything new happen in this regime?

\section*{Acknowledgements}
We thank Veronika Hubeny and Mukund Rangamani for useful discussions and
comments on this work, and Netta Engelhardt, Gary Horowitz, Juan Maldacena,
Don Marolf, and Ed Witten for earlier relevant discussions. This work was supported by the U.S. Department of Energy grant
DE-SC0009999. 

\appendix
\section{Appendix}\label{sec:appendix}
\subsection{Coefficients}\label{sec:coefs}
For conciseness, we denote $\Delta_{-}=\Delta$ in this subsection.
Then the coefficients in \eqref{eq:phifalloff} read \cite{HenMar06}
\begin{equation}
\begin{aligned}
    b_1 &= \frac{ g_3 }{ \Delta(\Delta - 1) }, \\
    b_2 &= \frac{ \Delta(3-2\Delta) }{ 4(4\Delta-3)  } + \frac{ 2g_4
    }{ \Delta (4\Delta -3) } + \frac{ 3g_3^2 }{ \Delta^2 (\Delta
    -1)(4\Delta - 3) }, \\
    b_3 &= \frac{ 5g_5 }{ 3\Delta (5\Delta - 3) }+\frac{ 4g_3 g_4 (5\Delta -
    4) }{ \Delta^2 (\Delta-1)(4\Delta - 3)(5\Delta - 3) }, \\
    &\qquad + \frac{
        g_3^3(10\Delta - 9) }{ \Delta^3 (5\Delta -3)(4\Delta-3)(\Delta
    - 1)^2 } + \frac{ g_3(-153+327\Delta - 170\Delta^2) }{ 18(\Delta -
    1)(4\Delta - 3)(5\Delta - 3) }.
\end{aligned}
\end{equation}
The coefficients in \eqref{eq:mdiv} are given by
\begin{equation}
\begin{aligned}
c_1 = a_1, \quad c_2 = a_2, \quad c_3 = a_3 - a_1^2, \quad c_4 = a_4 - 2a_1 a_2,
\end{aligned}
\end{equation}
where
\begin{equation}
\begin{aligned}
    a_1 &= - \frac{ \Delta }{ 2 }, \\
    a_2 &= -\frac{ 4 }{ 3 }\Delta b_1, \\
    a_3 &= - \frac{ \Delta }{ 4  }\left(-\frac{ \Delta }{ 2 } +
    6b_{2}+4b_{1}^2\right), \\
    a_4 &= -\frac{ \Delta }{ 5 }\left(8b_3 + 12b_1 b_2 - \frac{ 10 }{ 3
    }\Delta b_1 \right).
\end{aligned}
\end{equation}

\subsection{The perturbative superpotential}\label{sec:superpotapp}
Solving perturbatively using an analytic ansatz for $P_-$, we find 
\begin{equation}\label{eq:P0expr}
\begin{aligned}
    P_{-} = 1 + \frac{ \Delta_{-} }{ 4 }\phi^2 + \frac{ g_3 }{ 6(\Delta_{-}-1)
    }\phi^3 + \frac{ -8g_3^3 + (\Delta_- - 1)^2(3\Delta_-^2 + 16 g_4) }{
        32(\Delta_{-} - 1)^2(4\Delta_- -3) }\phi^4 + \ldots.
\end{aligned}
\end{equation}
To determine the non-analytic part, we expand $P_{-} = P_0(\phi) + \sum_i
s^{i}P_i(\phi)$ and demand that
\begin{equation}
\begin{aligned}
    \delta\left[2(P')^2 - 3P^2\right] = 0.
\end{aligned}
\end{equation}
The first order equation becomes
\begin{equation}
\begin{aligned}
\frac{ P'_1 }{ P_1 } = \frac{ 3 }{ 2 } \frac{ P_0 }{ P'_0 }
\end{aligned}
\end{equation}
Thus 
\begin{equation}
\begin{aligned}
\ln |P_1| = c + \frac{ 3 }{ 2 }\int \dd \phi \frac{ P_{0} }{ P_0' }
\end{aligned}
\end{equation}
Now define the regular quantity
\begin{equation}
\begin{aligned}
    \gamma(\phi) \equiv \frac{ P_0 }{ P_0' } - \frac{ 2 }{ \Delta_{-} \phi } =
    \mathcal{O}(1)
\end{aligned}
\end{equation}
Then we find that
\begin{equation}
\begin{aligned}
    P_{1} = C|\phi|^\frac{ 3 }{ \Delta_{-} } T(\phi)
\end{aligned}
\end{equation}
for a constant $C$, and with
\begin{equation}
\begin{aligned}
    T(\phi) = e^{\frac{ 3 }{ 2 }\int_{0}^{\phi}\dd \phi' \gamma(\phi')} = 1 +
    \mathcal{O}(\phi) 
\end{aligned}
\end{equation}
manifestly analytic. Note that there is another non-analytic branch,
since when removing the absolute value on $P_1$, we can decide whether to include
a $\sign\phi$ term. This makes $P_1$ anti-symmetric instead of symmetric near
$\phi=0$
to leading order. To determine $s_c$, we always want the symmetric branch, since the
anti-symmetric branch gives no lower bound on the mass when we have conformal
boundary conditions (in this case, we get an
extra $\sign(\alpha)$ factor in the term involving $s_c$ in \eqref{eq:Wbndconf}).

Next, going to second order we find the equation
\begin{equation}
\begin{aligned}
P_2' - \frac{ 3 P_0 }{ 2 P_{0}'}P_2 = 
    \frac{ 3 }{ 4 }\frac{ P_1^2 }{ P_0'  }-\frac{ (P_1')^2 }{ 2P_0' }
\end{aligned}
\end{equation}
This equation is solved with an integrating factor. Setting
\begin{equation}
\begin{aligned}
    P_2 = G(\phi) e^{\int_{0}^{\phi}\dd \phi' \frac{ 3P_0(\phi') }{ 2P_{0}(\phi')
    }} = G(\phi) T(\phi)|\phi|^{3/\Delta_{-}},
\end{aligned}
\end{equation}
we get
\begin{equation}
\begin{aligned}
    G'(\phi) &\equiv \frac{ 1 }{ T(\phi)|\phi|^{3/\Delta_{-}} } \frac{ 1 }{ 2P_{0}'(\phi) }\left(\frac{
        3}{ 2 }P_{1}(\phi)^2 - P'_1(\phi)^2 \right).
\end{aligned}
\end{equation}
The homogeneous part is just a shift to $P_1$, which we can conventionally set to zero.
Plugging in $P_1$ and integrating we find to leading order
\begin{equation}
\begin{aligned}
    P_2 = C^2 |\phi|^{6/\Delta_{-} - 2}\frac{ 9 }{ \Delta_{-}^2(2\Delta_{-} -3)
    }.
\end{aligned}
\end{equation}

\subsection{Determining $s_c$}\label{sec:sc}
The precise value of $s_c$ is important to us. An
imprecise determination of $s_c$ can lead us to falsely conclude the PI is violated
in a PMT-respecting theory. We will determine $s_c$ in two independent ways. 

\subsubsection*{Direct determination of $s_c$}
The first way is the direct way, where we numerically solve \eqref{eq:VWpot}
for $P$ to determine if (1) $P_-$ exists for any $s$, and (2) if it does, what is
the value of $s_c$.
First, we rewrite \eqref{eq:VWpot} as 
\begin{equation}
\begin{aligned}
    P'(\phi) = \pm \sqrt{\frac{ 1 }{ 4 } V(\phi) + \frac{ 3 }{ 4 }P(\phi)^2 },
\end{aligned}
\end{equation}
where the plus (minus) sign is chosen for $\phi>0$ ($\phi<0$). This ODE is
singular at $\phi=0$, so we cannot numerically integrate it from $\phi=0$. Instead, we
must solve using a series expansion near $\phi=0$ and then integrate numerically it from
$\phi=\pm \epsilon$ to larger and smaller $\phi$. We use the series expansion to set the initial
condition for $P'(\pm \epsilon)$. 

Now, the series solution for $P_-$ near $\phi=0$ takes the form of a double
perturbative series in $\phi$ and $s$:
\begin{equation}
\begin{aligned}
    P_{-}(\phi;s) = P_{0}(\phi) + s P_{1}(\phi) + s^2 P_2(\phi) + \ldots,
\end{aligned}
\end{equation}
where the analytic part $P_0$ is given by \eqref{eq:P0expr}.
Since we want to study
possibly large values of $s$, it looks concerning that we are working with a
series expansion in $s$. However, the leading term in $P_i$ is
proportional to $|\phi|^{\gamma_i}$ where the exponent $\gamma_i$ increases with
$i$, so at $\phi=\epsilon$, the ratio between two successive terms scales like
$s|\epsilon|^{\gamma_{i+1}-\gamma_{i}}$, which for any $s$ is small for sufficiently small
$\epsilon$. 

The coefficients $\gamma_{i}$ all satisfy $\gamma_{i}> 2$, but in the limit
$\Delta_{-}\rightarrow 3/2$ where we approach BF bound saturation, all
$\gamma_i$ approach $2$ \cite{AmsRob11}. Thus, close to $\Delta_{-}=3/2$ the
series converges very slowly. We will never go very close to this value: we
always consider $\Delta_{-}\leq 5/4$. 

In picking $\epsilon$, we must not pick it too small or too large. If
$\epsilon$ is too small, the contribution to $P'(\pm\epsilon)$ from $s
P_1(\epsilon)\sim s|\epsilon|^{3/\Delta_{-}}$ competes with numerical noise,
so we do not get reliable
results. On the other hand, we cannot have too large $\epsilon$ either, as this
breaks the
perturbative treatment. In practice, for a given pair $\epsilon$ and $s$, we are
satisfied as long as $\epsilon$ and the ratio $s P_2(\epsilon)/P_1(\epsilon)\sim
s|\epsilon|^{3/\Delta_{-}-2}$
is small. In practice $\epsilon\in [10^{-2}, 10^{-3}]$ works well. 
See Appendix \ref{sec:superpotapp} the leading-orderexpressions of $P_1$ and
$P_2$.

\subsubsection*{Indirect determination of $s_c$}
In \cite{FauHor10}, they found an alternative way to extract $s_c$. First, they
used the fact that for theories with a superpotential $P$, there is a one-to-one
correspondence between superpotentials and planar domain
walls \cite{DeWFree00,FreNun04}. The latter are planar-symmetric stationary
solutions 
\begin{equation}\label{eq:domainwall}
\begin{aligned}
    \dd s^2 = -f(r)\dd t^2 + \frac{ \dd r^2 }{ g(r)^2 } + r^2(\dd x^2 + \dd
    y^2).
\end{aligned}
\end{equation}
Focusing on $P_-$, we have the following: a superpotential $P_{-}(\phi;s)$ corresponds to a one-parameter family of domain
walls, whose individual members are related by a scaling symmetry. Each member of
the family satisfies $\beta = -s\sign\alpha |\alpha|^{\frac{ \Delta_{+} }{ \Delta_{-}
}}$. However, only the family corresponding to $P_{-}(\phi; s_c)$ is regular
near $r=0$.  

Next, consider spherical solitons, which have the same metric as \eqref{eq:domainwall}
except that we replace $\dd x^2 + \dd y^2 \rightarrow \dd \Omega^2$.
It is reasonable to expect that these approach the regular domain walls in the
high energy limit ($|\alpha|\rightarrow \infty$). Thus, if we compute the
solitons numerically, we can extract $s_c$ by fitting $\beta_{\rm soliton}(\alpha)$ to the
expression $\beta = s\sign\alpha |\alpha|^{\frac{ \Delta_{+} }{ \Delta_{-}
}}$ at high energies. We will refer to this as the soliton method of extracting
$s_c$. We will not explain how to construct the solitons, since this has been
explained in the literature many times -- see for example \cite{HerHor04b}. 

There is one remaining puzzle here, which was not discussed in \cite{FauHor10}. If $V(\phi)$ is not symmetric, we
find different values for $s_c$ depending on whether we send $\alpha \rightarrow
+\infty$ or $\alpha \rightarrow -\infty$. The solution to this puzzle is the
following: there in fact exist two non-analytic branches for $P_{-}$. Rather than having
the leading nonanalytic behavior go as $|\phi|^{3/\Delta_{-}}$, we can have it
be $\sign \phi |\phi|^{3/\Delta_{-}}$. One of the asymptotic regimes gives an
$s$-value corresponding to the critical $s$ for the $P_-$ branch corresponding
to $\sign \phi |\phi|^{3/\Delta_{-}}$. This branch gives no lower bound on the
mass however. We can determine which $s$ is correct by comparing to the direct
method. We always find that the smallest value of the two $s$-values obtained
from solitons correspond to $s_c$.

\subsection{Numerics}\label{sec:numerics}
To solve \eqref{eq:maineq} and \eqref{eq:maineq2}, we solve for
the $\mathcal{O}(1)$ variables $(\hat{\alpha}(r), h(r), \Gamma(r))$ where
\begin{equation}
\begin{aligned}
    \hat{\alpha}(r) &\equiv r^{\Delta_{-}}\phi(r), \\
    h(r) &\equiv \frac{ 1 }{ r^3 }H(r).
\end{aligned}
\end{equation}
The function $m(r)$ can be computed as
\begin{equation}
\begin{aligned}
    m(r) = r^{3} + \frac{ 1 }{ \Gamma(r) }(r_* + r^3 h(r)).
\end{aligned}
\end{equation}
We use Mathematica's built-in NDSolve method with an explicit fourth-order Runge-Kutta scheme.
We find that fourth-order RK yields better (less noisy) solutions at large $r$ than other
methods. We impose a maximum step size for $\Delta r=1$ in the integration and
integrate to a maximal $r$ of $r_{\rm max}$ ranging from $10^4$ to $5\times 10^5$. For
most scaling dimensions $r_{\rm max} \sim 10^4$ is more than sufficient. However, when we get
close to $\Delta = 1/2$, specifically in Sec.~\ref{sec:th6}, quantities of
interest converge slowly, and our extraction of
$\beta$ starts to converge roughly around $r_{\rm max} \sim [2, 8] \times 10^{4}$. 
However, in the upper parts of this range, our determination of $M_0$ becomes 
noisy. As a compromise we work with $r_{\rm max} \sim 4 \times 10^{4}$ in
this case. It would be good to do a more careful study in this regime, but we do
find results consistent with other scaling dimensions and the proven PMTs with
out current numerics.

\bibliographystyle{jhep}
\bibliography{all}

\providecommand{\href}[2]{#2}\begingroup\raggedright\begin{thebibliography}{10}

\bibitem{Wil12}
F.~Wilczek, {\it {Quantum Time Crystals}},  {\em Phys. Rev. Lett.} {\bf 109} (2012) 160401, [\href{http://arxiv.org/abs/1202.2539}{{\tt arXiv:1202.2539}}].

\bibitem{Bru12}
P.~Bruno, {\it {Comment on \textquotedblleft{}Quantum Time Crystals\textquotedblright{}}},  {\em Phys. Rev. Lett.} {\bf 110} (2013), no.~11 118901, [\href{http://arxiv.org/abs/1210.4128}{{\tt arXiv:1210.4128}}].

\bibitem{Wil13}
F.~Wilczek, {\it Wilczek reply:},  {\em Phys. Rev. Lett.} {\bf 110} (Mar, 2013) 118902.

\bibitem{LiGon12}
T.~Li, Z.-X. Gong, Z.-Q. Yin, H.~T. Quan, X.~Yin, P.~Zhang, L.-M. Duan, and X.~Zhang, {\it Space-time crystals of trapped ions},  {\em Phys. Rev. Lett.} {\bf 109} (Oct, 2012) 163001.

\bibitem{Bru13}
P.~Bruno, {\it Comment on ``space-time crystals of trapped ions''},  {\em Phys. Rev. Lett.} {\bf 111} (Jul, 2013) 029301.

\bibitem{TonZhe13}
T.~Li, Z.-X. Gong, Z.-Q. Yin, H.~T. Quan, X.~Yin, P.~Zhang, L.~M. Duan, and X.~Zhang, {\it Reply to comment on "space-time crystals of trapped ions"},  2013.

\bibitem{Bru13b}
P.~Bruno, {\it Impossibility of spontaneously rotating time crystals: A no-go theorem},  {\em Phys. Rev. Lett.} {\bf 111} (Aug, 2013) 070402.

\bibitem{WatOsh14}
H.~Watanabe and M.~Oshikawa, {\it {Absence of Quantum Time Crystals}},  {\em Phys. Rev. Lett.} {\bf 114} (2015), no.~25 251603, [\href{http://arxiv.org/abs/1410.2143}{{\tt arXiv:1410.2143}}].

\bibitem{Sac14}
K.~Sacha, {\it {Modeling spontaneous breaking of time-translation symmetry}},  {\em Phys. Rev. A} {\bf 91} (2015), no.~3 033617, [\href{http://arxiv.org/abs/1410.3638}{{\tt arXiv:1410.3638}}].

\bibitem{ElsBau16}
D.~V. Else, B.~Bauer, and C.~Nayak, {\it {Floquet Time Crystals}},  {\em Phys. Rev. Lett.} {\bf 117} (2016), no.~9 090402.

\bibitem{Els16}
D.~V. Else, B.~Bauer, and C.~Nayak, {\it {Prethermal Phases of Matter Protected by Time-Translation Symmetry}},  {\em Phys. Rev. X} {\bf 7} (2017), no.~1 011026, [\href{http://arxiv.org/abs/1607.05277}{{\tt arXiv:1607.05277}}].

\bibitem{Zha17}
J.~Zhang et~al., {\it {Observation of a discrete time crystal}},  {\em Nature} {\bf 543} (2017) 217--220.

\bibitem{KheLaz16}
V.~Khemani, A.~Lazarides, R.~Moessner, and L.~Sondhi, S.~\, {\it {Phase Structure of Driven Quantum Systems}},  {\em Phys. Rev. Lett.} {\bf 116} (2016), no.~25 250401.

\bibitem{Cho17}
S.~Choi et~al., {\it {Observation of discrete time-crystalline order in a disordered dipolar many-body system}},  {\em Nature} {\bf 543} (2017), no.~7644 221--225.

\bibitem{YaoPot17}
N.~Y. Yao, A.~C. Potter, I.-D. Potirniche, and A.~Vishwanath, {\it Discrete time crystals: Rigidity, criticality, and realizations},  {\em Phys. Rev. Lett.} {\bf 118} (Jan, 2017) 030401.

\bibitem{KheMoe19}
V.~Khemani, R.~Moessner, and S.~L. Sondhi, {\it {A Brief History of Time Crystals}},  \href{http://arxiv.org/abs/1910.10745}{{\tt arXiv:1910.10745}}.

\bibitem{tHo74}
G.~'t~Hooft, {\it {A Planar Diagram Theory for Strong Interactions}},  {\em Nucl. Phys. B} {\bf 72} (1974) 461.

\bibitem{Mal97}
J.~Maldacena, {\it The large {$N$} limit of superconformal field theories and supergravity},  {\em Adv. Theor. Math. Phys.} {\bf 2} (1998) 231, [\href{http://arxiv.org/abs/hep-th/9711200}{{\tt hep-th/9711200}}].

\bibitem{FigRos23}
P.~Figueras and L.~Rossi, {\it {Non-linear instability of slowly rotating Kerr-AdS black holes}},  \href{http://arxiv.org/abs/2311.14167}{{\tt arXiv:2311.14167}}.

\bibitem{GauSon09}
J.~P. Gauntlett, J.~Sonner, and T.~Wiseman, {\it {Holographic superconductivity in M-Theory}},  {\em Phys. Rev. Lett.} {\bf 103} (2009) 151601, [\href{http://arxiv.org/abs/0907.3796}{{\tt arXiv:0907.3796}}].

\bibitem{Her06}
T.~Hertog, {\it {Towards a Novel no-hair Theorem for Black Holes}},  {\em Phys. Rev. D} {\bf 74} (2006) 084008, [\href{http://arxiv.org/abs/gr-qc/0608075}{{\tt gr-qc/0608075}}].

\bibitem{Fol22}
{\AA}.~Folkestad, {\it {Penrose Inequality as a Constraint on the Low Energy Limit of Quantum Gravity}},  {\em Phys. Rev. Lett.} {\bf 130} (2023), no.~12 121501, [\href{http://arxiv.org/abs/2209.00013}{{\tt arXiv:2209.00013}}].

\bibitem{Vaf05}
C.~Vafa, {\it The string landscape and the swampland},  \href{http://arxiv.org/abs/hep-th/0509212}{{\tt hep-th/0509212}}.

\bibitem{OogVaf06}
H.~Ooguri and C.~Vafa, {\it On the geometry of the string landscape and the swampland},  {\em Nucl. Phys.} {\bf B766} (2007) 21--33, [\href{http://arxiv.org/abs/hep-th/0605264}{{\tt hep-th/0605264}}].

\bibitem{Mal01}
J.~M. Maldacena, {\it {Eternal black holes in anti-de Sitter}},  {\em JHEP} {\bf 04} (2003) 021, [\href{http://arxiv.org/abs/hep-th/0106112}{{\tt hep-th/0106112}}].

\bibitem{HawPag83}
S.~Hawking and D.~N. Page, {\it {Thermodynamics of Black Holes in anti-De Sitter Space}},  {\em Commun.Math.Phys.} {\bf 87} (1983) 577.

\bibitem{Mar18}
D.~Marolf, {\it {Microcanonical Path Integrals and the Holography of small Black Hole Interiors}},  {\em JHEP} {\bf 09} (2018) 114, [\href{http://arxiv.org/abs/1808.00394}{{\tt arXiv:1808.00394}}].

\bibitem{KunLuc06}
H.~K. Kunduri, J.~Lucietti, and H.~S. Reall, {\it {Gravitational perturbations of higher dimensional rotating black holes: Tensor perturbations}},  {\em Phys. Rev. D} {\bf 74} (2006) 084021, [\href{http://arxiv.org/abs/hep-th/0606076}{{\tt hep-th/0606076}}].

\bibitem{DiaHor11}
O.~J.~C. Dias, G.~T. Horowitz, and J.~E. Santos, {\it {Black holes with only one Killing field}},  {\em JHEP} {\bf 07} (2011) 115, [\href{http://arxiv.org/abs/1105.4167}{{\tt arXiv:1105.4167}}].

\bibitem{DiaSan15}
O.~J.~C. Dias, J.~E. Santos, and B.~Way, {\it {Black holes with a single Killing vector field: black resonators}},  {\em JHEP} {\bf 12} (2015) 171, [\href{http://arxiv.org/abs/1505.04793}{{\tt arXiv:1505.04793}}].

\bibitem{IshMur18}
T.~Ishii and K.~Murata, {\it {Black resonators and geons in AdS5}},  {\em Class. Quant. Grav.} {\bf 36} (2019), no.~12 125011, [\href{http://arxiv.org/abs/1810.11089}{{\tt arXiv:1810.11089}}].

\bibitem{IshMur21}
T.~Ishii, K.~Murata, J.~E. Santos, and B.~Way, {\it {Multioscillating black holes}},  {\em JHEP} {\bf 05} (2021) 011, [\href{http://arxiv.org/abs/2101.06325}{{\tt arXiv:2101.06325}}].

\bibitem{Ish21}
T.~Ishii, {\it {Superradiant instability and black resonators in AdS}},  {\em SciPost Phys. Proc.} {\bf 4} (2021) 008.

\bibitem{HawRea00}
S.~W. Hawking and H.~S. Reall, {\it {Charged and rotating AdS black holes and their CFT duals}},  {\em Phys. Rev. D} {\bf 61} (2000) 024014, [\href{http://arxiv.org/abs/hep-th/9908109}{{\tt hep-th/9908109}}].

\bibitem{CarDia04}
V.~Cardoso and O.~J.~C. Dias, {\it {Small Kerr-anti-de Sitter black holes are unstable}},  {\em Phys. Rev. D} {\bf 70} (2004) 084011, [\href{http://arxiv.org/abs/hep-th/0405006}{{\tt hep-th/0405006}}].

\bibitem{CarDia13}
V.~Cardoso, O.~J.~C. Dias, G.~S. Hartnett, L.~Lehner, and J.~E. Santos, {\it {Holographic thermalization, quasinormal modes and superradiance in Kerr-AdS}},  {\em JHEP} {\bf 04} (2014) 183, [\href{http://arxiv.org/abs/1312.5323}{{\tt arXiv:1312.5323}}].

\bibitem{Zel71}
Y.~B. {Zel'Dovich}, {\it {Generation of Waves by a Rotating Body}},  {\em Soviet Journal of Experimental and Theoretical Physics Letters} {\bf 14} (Aug., 1971) 180.

\bibitem{Sta73}
A.~A. Starobinsky, {\it {Amplification of waves reflected from a rotating ''black hole''.}},  {\em Sov. Phys. JETP} {\bf 37} (1973), no.~1 28--32.

\bibitem{BriCar15}
R.~Brito, V.~Cardoso, and P.~Pani, {\it {Superradiance}: {New Frontiers in Black Hole Physics}},  {\em Lect. Notes Phys.} {\bf 906} (2015) pp.1--237, [\href{http://arxiv.org/abs/1501.06570}{{\tt arXiv:1501.06570}}].

\bibitem{PreTeu72}
W.~H. Press and S.~A. Teukolsky, {\it {Floating Orbits, Superradiant Scattering and the Black-hole Bomb}},  {\em Nature} {\bf 238} (1972) 211--212.

\bibitem{DiaHor11b}
O.~J.~C. Dias, G.~T. Horowitz, and J.~E. Santos, {\it {Gravitational Turbulent Instability of Anti-de Sitter Space}},  {\em Class. Quant. Grav.} {\bf 29} (2012) 194002, [\href{http://arxiv.org/abs/1109.1825}{{\tt arXiv:1109.1825}}].

\bibitem{HorSan14}
G.~T. Horowitz and J.~E. Santos, {\it {Geons and the instability of anti-de Sitter spacetime}},  {\em Surveys Diff. Geom.} {\bf 20} (2015), no.~1 321--335, [\href{http://arxiv.org/abs/1408.5906}{{\tt arXiv:1408.5906}}].

\bibitem{MarFod17}
G.~Martinon, G.~Fodor, P.~Grandcl\'ement, and P.~Forg\`acs, {\it {Gravitational geons in asymptotically anti-de Sitter spacetimes}},  {\em Class. Quant. Grav.} {\bf 34} (2017), no.~12 125012, [\href{http://arxiv.org/abs/1701.09100}{{\tt arXiv:1701.09100}}].

\bibitem{GreHol15}
S.~R. Green, S.~Hollands, A.~Ishibashi, and R.~M. Wald, {\it {Superradiant instabilities of asymptotically anti-de Sitter black holes}},  {\em Class. Quant. Grav.} {\bf 33} (2016), no.~12 125022, [\href{http://arxiv.org/abs/1512.02644}{{\tt arXiv:1512.02644}}].

\bibitem{CheLow18}
P.~M. Chesler and D.~A. Lowe, {\it {Nonlinear Evolution of the AdS$_4$ Superradiant Instability}},  {\em Phys. Rev. Lett.} {\bf 122} (2019), no.~18 181101, [\href{http://arxiv.org/abs/1801.09711}{{\tt arXiv:1801.09711}}].

\bibitem{Che21}
P.~M. Chesler, {\it {Hairy black resonators and the AdS4 superradiant instability}},  {\em Phys. Rev. D} {\bf 105} (2022), no.~2 024026, [\href{http://arxiv.org/abs/2109.06901}{{\tt arXiv:2109.06901}}].

\bibitem{KimKun23}
S.~Kim, S.~Kundu, E.~Lee, J.~Lee, S.~Minwalla, and C.~Patel, {\it {Grey Galaxies\textquoteright{} as an endpoint of the Kerr-AdS superradiant instability}},  {\em JHEP} {\bf 11} (2023) 024, [\href{http://arxiv.org/abs/2305.08922}{{\tt arXiv:2305.08922}}].

\bibitem{HolSmu11}
G.~Holzegel and J.~Smulevici, {\it {Decay properties of Klein-Gordon fields on Kerr-AdS spacetimes}},  {\em Commun. Pure Appl. Math.} {\bf 66} (2013) 1751--1802, [\href{http://arxiv.org/abs/1110.6794}{{\tt arXiv:1110.6794}}].

\bibitem{HolWar14}
G.~H. Holzegel and C.~M. Warnick, {\it {Boundedness and growth for the massive wave equation on asymptotically anti-de Sitter black holes}},  {\em J. Funct. Anal.} {\bf 266} (2014), no.~4 2436--2485, [\href{http://arxiv.org/abs/1209.3308}{{\tt arXiv:1209.3308}}].

\bibitem{Pen73}
R.~{Penrose}, {\it {Naked Singularities}},  in {\em Sixth Texas Symposium on Relativistic Astrophysics} (D.~J. {Hegyi}, ed.), vol.~224 of {\em Annals of the New York Academy of Sciences}, p.~125, 1973.

\bibitem{Pen65}
R.~Penrose, {\it Gravitational collapse and space-time singularities},  {\em Phys. Rev. Lett.} {\bf 14} (Jan, 1965) 57--59.

\bibitem{HawEll}
S.~W. Hawking and G.~F.~R. Ellis, {\em The large scale stucture of space-time}.
\newblock Cambridge University Press, Cambridge, England, 1973.

\bibitem{Wald}
R.~M. Wald, {\em General Relativity}.
\newblock The University of Chicago Press, Chicago, 1984.

\bibitem{Haw71}
S.~W. Hawking, {\it Gravitational radiation from colliding black holes},  {\em Phys. Rev. Lett.} {\bf 26} (1971) 1344--1346.

\bibitem{Ger73}
R.~Geroch, {\it Energy extraction*},  {\em Annals of the New York Academy of Sciences} {\bf 224} (1973), no.~1 108--117, [\href{http://arxiv.org/abs/https://nyaspubs.onlinelibrary.wiley.com/doi/pdf/10.1111/j.1749-6632.1973.tb41445.x}{{\tt https://nyaspubs.onlinelibrary.wiley.com/doi/pdf/10.1111/j.1749-6632.1973.tb41445.x}}].

\bibitem{HuiIlm01}
G.~Huisken and T.~Ilmanen, {\it {The Inverse Mean Curvature Flow and the Riemannian Penrose Inequality}},  {\em Journal of Differential Geometry} {\bf 59} (2001), no.~3 353 -- 437.

\bibitem{Bra99}
H.~L. {Bray}, {\it {Proof of the Riemannian Penrose Conjecture Using the Positive Mass Theorem}},  {\em arXiv Mathematics e-prints} (Nov., 1999) math/9911173, [\href{http://arxiv.org/abs/math/9911173}{{\tt math/9911173}}].

\bibitem{Bra07}
H.~L. Bray and D.~A. Lee, {\it {On the Riemannian Penrose inequality in dimensions less than 8}},  {\em Duke Math. J.} {\bf 148} (2009) 81--106, [\href{http://arxiv.org/abs/0705.1128}{{\tt arXiv:0705.1128}}].

\bibitem{LevFre12}
L.~L. de~Lima and F.~Girao, {\it Positive mass and penrose type inequalities for asymptotically hyperbolic hypersurfaces},  2012.

\bibitem{GeGuo13}
Y.~Ge, G.~Wang, J.~Wu, and C.~Xia, {\it A penrose inequality for graphs over kottler space},  2013.

\bibitem{HusViq17}
V.~Husain and S.~Singh, {\it {Penrose inequality in anti\textendash{}de Sitter space}},  {\em Phys. Rev. D} {\bf 96} (2017), no.~10 104055, [\href{http://arxiv.org/abs/1709.02395}{{\tt arXiv:1709.02395}}].

\bibitem{EngHor19}
N.~Engelhardt and G.~T. Horowitz, {\it {Holographic argument for the Penrose inequality in AdS spacetimes}},  {\em Phys. Rev. D} {\bf 99} (2019), no.~12 126009, [\href{http://arxiv.org/abs/1903.00555}{{\tt arXiv:1903.00555}}].

\bibitem{Schw06}
F.~Schwabl, {\em Statistical Mechanics}.
\newblock Springer Berlin Heidelberg, 2nd ed. 2006~ed., 2006.

\bibitem{KhuWei14}
M.~Khuri, G.~Weinstein, and S.~Yamada, {\it {Extensions of the Charged Riemannian Penrose Inequality}},  {\em Class. Quant. Grav.} {\bf 32} (2015), no.~3 035019, [\href{http://arxiv.org/abs/1410.5027}{{\tt arXiv:1410.5027}}].

\bibitem{HerHor03}
T.~Hertog, G.~T. Horowitz, and K.~Maeda, {\it {Negative energy density in Calabi-Yau compactifications}},  {\em JHEP} {\bf 05} (2003) 060, [\href{http://arxiv.org/abs/hep-th/0304199}{{\tt hep-th/0304199}}].

\bibitem{BreFre82}
P.~Breitenlohner and D.~Z. Freedman, {\it Positive energy in anti-de {S}itter backgrounds {AND} gauged extended supergravity},  {\em Phys. Lett.} {\bf B115} (1982) 197.

\bibitem{Har08}
S.~A. Hartnoll, C.~P. Herzog, and G.~T. Horowitz, {\it {Holographic Superconductors}},  {\em JHEP} {\bf 12} (2008) 015, [\href{http://arxiv.org/abs/0810.1563}{{\tt arXiv:0810.1563}}].

\bibitem{Har08b}
S.~A. Hartnoll, C.~P. Herzog, and G.~T. Horowitz, {\it {Building a Holographic Superconductor}},  {\em Phys. Rev. Lett.} {\bf 101} (2008) 031601, [\href{http://arxiv.org/abs/0803.3295}{{\tt arXiv:0803.3295}}].

\bibitem{SudGon02}
D.~Sudarsky and J.~A. Gonzalez, {\it {On black hole scalar hair in asymptotically anti-de Sitter space-times}},  {\em Phys. Rev. D} {\bf 67} (2003) 024038, [\href{http://arxiv.org/abs/gr-qc/0207069}{{\tt gr-qc/0207069}}].

\bibitem{TorMae01}
T.~Torii, K.~Maeda, and M.~Narita, {\it Scalar hair on the black hole in asymptotically anti--de sitter spacetime},  {\em Phys. Rev. D} {\bf 64} (Jul, 2001) 044007.

\bibitem{MarTro04}
C.~Martinez, R.~Troncoso, and J.~Zanelli, {\it {Exact black hole solution with a minimally coupled scalar field}},  {\em Phys. Rev. D} {\bf 70} (2004) 084035, [\href{http://arxiv.org/abs/hep-th/0406111}{{\tt hep-th/0406111}}].

\bibitem{Zlo04}
K.~G. Zloshchastiev, {\it {On co-existence of black holes and scalar field}},  {\em Phys. Rev. Lett.} {\bf 94} (2005) 121101, [\href{http://arxiv.org/abs/hep-th/0408163}{{\tt hep-th/0408163}}].

\bibitem{HerMae04}
T.~Hertog and K.~Maeda, {\it {Black holes with scalar hair and asymptotics in N = 8 supergravity}},  {\em JHEP} {\bf 07} (2004) 051, [\href{http://arxiv.org/abs/hep-th/0404261}{{\tt hep-th/0404261}}].

\bibitem{HenMar06}
M.~Henneaux, C.~Martinez, R.~Troncoso, and J.~Zanelli, {\it {Asymptotic behavior and Hamiltonian analysis of anti-de Sitter gravity coupled to scalar fields}},  {\em Annals Phys.} {\bf 322} (2007) 824--848, [\href{http://arxiv.org/abs/hep-th/0603185}{{\tt hep-th/0603185}}].

\bibitem{HerHor04b}
T.~Hertog and G.~T. Horowitz, {\it {Designer gravity and field theory effective potentials}},  {\em Phys. Rev. Lett.} {\bf 94} (2005) 221301, [\href{http://arxiv.org/abs/hep-th/0412169}{{\tt hep-th/0412169}}].

\bibitem{Wit01b}
E.~Witten, {\it {Multitrace operators, boundary conditions, and AdS / CFT correspondence}},  \href{http://arxiv.org/abs/hep-th/0112258}{{\tt hep-th/0112258}}.

\bibitem{HerHol05}
T.~Hertog and S.~Hollands, {\it {Stability in designer gravity}},  {\em Class. Quant. Grav.} {\bf 22} (2005) 5323--5342, [\href{http://arxiv.org/abs/hep-th/0508181}{{\tt hep-th/0508181}}].

\bibitem{AmsMar06}
A.~J. Amsel and D.~Marolf, {\it {Energy Bounds in Designer Gravity}},  {\em Phys. Rev. D} {\bf 74} (2006) 064006, [\href{http://arxiv.org/abs/hep-th/0605101}{{\tt hep-th/0605101}}]. [Erratum: Phys.Rev.D 75, 029901 (2007)].

\bibitem{AmsHer07}
A.~J. Amsel, T.~Hertog, S.~Hollands, and D.~Marolf, {\it {A Tale of two superpotentials: Stability and instability in designer gravity}},  {\em Phys. Rev. D} {\bf 75} (2007) 084008, [\href{http://arxiv.org/abs/hep-th/0701038}{{\tt hep-th/0701038}}]. [Erratum: Phys.Rev.D 77, 049903 (2008)].

\bibitem{Ams12}
A.~J. Amsel and M.~M. Roberts, {\it {Stability in Holographic Theories with Irrelevant Deformations}},  {\em Phys. Rev. D} {\bf 87} (2013) 086007, [\href{http://arxiv.org/abs/1211.2840}{{\tt arXiv:1211.2840}}].

\bibitem{Wit81}
E.~Witten, {\it {A Simple Proof of the Positive Energy Theorem}},  {\em Commun. Math. Phys.} {\bf 80} (1981) 381.

\bibitem{Nes81}
J.~A. Nester, {\it {A New gravitational energy expression with a simple positivity proof}},  {\em Phys. Lett. A} {\bf 83} (1981) 241.

\bibitem{Bou84}
W.~Boucher, {\it Positive energy without supersymmetry},  {\em Nucl. Phys.} {\bf B242} (1984) 282.

\bibitem{Tow84}
P.~K. Townsend, {\it {Positive Energy and the Scalar Potential in Higher Dimensional (Super)gravity Theories}},  {\em Phys. Lett. B} {\bf 148} (1984) 55--59.

\bibitem{Pap06}
I.~Papadimitriou, {\it {Non-Supersymmetric Membrane Flows from Fake Supergravity and Multi-Trace Deformations}},  {\em JHEP} {\bf 02} (2007) 008, [\href{http://arxiv.org/abs/hep-th/0606038}{{\tt hep-th/0606038}}].

\bibitem{Pap07}
I.~Papadimitriou, {\it {Multi-Trace Deformations in AdS/CFT: Exploring the Vacuum Structure of the Deformed CFT}},  {\em JHEP} {\bf 05} (2007) 075, [\href{http://arxiv.org/abs/hep-th/0703152}{{\tt hep-th/0703152}}].

\bibitem{FauHor10}
T.~Faulkner, G.~T. Horowitz, and M.~M. Roberts, {\it {New stability results for Einstein scalar gravity}},  {\em Class. Quant. Grav.} {\bf 27} (2010) 205007, [\href{http://arxiv.org/abs/1006.2387}{{\tt arXiv:1006.2387}}].

\bibitem{EngFol21}
N.~Engelhardt and {\AA}.~Folkestad, {\it {General bounds on holographic complexity}},  {\em JHEP} {\bf 01} (2022) 040, [\href{http://arxiv.org/abs/2109.06883}{{\tt arXiv:2109.06883}}].

\bibitem{HerHor04c}
T.~Hertog, G.~T. Horowitz, and K.~Maeda, {\it {Generic cosmic censorship violation in anti-de Sitter space}},  {\em Phys. Rev. Lett.} {\bf 92} (2004) 131101, [\href{http://arxiv.org/abs/gr-qc/0307102}{{\tt gr-qc/0307102}}].

\bibitem{CveDuf99}
M.~Cvetic, M.~J. Duff, P.~Hoxha, J.~T. Liu, H.~Lu, J.~X. Lu, R.~Martinez-Acosta, C.~N. Pope, H.~Sati, and T.~A. Tran, {\it {Embedding AdS black holes in ten-dimensions and eleven-dimensions}},  {\em Nucl. Phys. B} {\bf 558} (1999) 96--126, [\href{http://arxiv.org/abs/hep-th/9903214}{{\tt hep-th/9903214}}].

\bibitem{DHoFre99}
E.~D'Hoker, D.~Z. Freedman, S.~D. Mathur, A.~Matusis, and L.~Rastelli, {\it {Extremal correlators in the AdS / CFT correspondence}},  \href{http://arxiv.org/abs/hep-th/9908160}{{\tt hep-th/9908160}}.

\bibitem{CasMar24}
A.~Castro and P.~J. Martinez, {\it {Revisiting extremal couplings in AdS/CFT}},  {\em JHEP} {\bf 12} (2024) 157, [\href{http://arxiv.org/abs/2409.15410}{{\tt arXiv:2409.15410}}].

\bibitem{AmsRob11}
A.~J. Amsel and M.~M. Roberts, {\it {Stability in Einstein-Scalar Gravity with a Logarithmic Branch}},  {\em Phys. Rev. D} {\bf 85} (2012) 106011, [\href{http://arxiv.org/abs/1112.3964}{{\tt arXiv:1112.3964}}].

\bibitem{DeWFree00}
O.~DeWolfe, D.~Z. Freedman, S.~S. Gubser, and A.~Karch, {\it {Modeling the fifth-dimension with scalars and gravity}},  {\em Phys. Rev. D} {\bf 62} (2000) 046008, [\href{http://arxiv.org/abs/hep-th/9909134}{{\tt hep-th/9909134}}].

\bibitem{FreNun04}
D.~Z. Freedman, C.~Nunez, M.~Schnabl, and K.~Skenderis, {\it {Fake supergravity and domain wall stability}},  {\em Phys. Rev. D} {\bf 69} (2004) 104027, [\href{http://arxiv.org/abs/hep-th/0312055}{{\tt hep-th/0312055}}].

\end{thebibliography}\endgroup

\end{document}